\documentclass[11pt,a4paper]{article}
\pdfoutput=1
\usepackage{jheppub}
\usepackage[english]{babel}
\usepackage{multirow}
\usepackage{color}
\usepackage[utf8]{inputenc}
\input{tcilatex}

\begin{document}

\begin{flushright}
PI/UAN-2017-617FT\\
LPT-Orsay-17-75
\end{flushright}

\title{Fermion Masses and Mixings and Dark Matter Constraints
in a Model with Radiative Seesaw Mechanism}
\author[1,2]{Nicol\'{a}s Bernal,}
\emailAdd{nicolas.bernal@uan.edu.co}
\author[3]{A. E. C\'{a}rcamo Hern\'{a}ndez,}
\emailAdd{antonio.carcamo@usm.cl}
\author[4]{Ivo de Medeiros Varzielas}
\emailAdd{ivo.de@udo.edu}
\author[3]{and Sergey Kovalenko}
\emailAdd{sergey.kovalenko@usm.cl}

\affiliation[1]{Centro de Investigaciones, Universidad Antonio Nariño\\
Carrera 3 Este \# 47A-15, Bogotá, Colombia}
\affiliation[2]{Laboratoire de Physique Théorique, CNRS,\\
Université Paris-Sud, Université Paris-Saclay, 91405 Orsay, France}
\affiliation[3]{Universidad T\'{e}cnica Federico Santa Mar\'{\i}a\\
and Centro Cient\'{\i}fico-Tecnol\'{o}gico de Valpara\'{\i}so\\
Casilla 110-V, Valpara\'{\i}so, Chile}
\affiliation[4]{CFTP, Departamento de F\'{\i}sica, Instituto Superior T\'{e}cnico,\\
Universidade de Lisboa, Avenida Rovisco Pais 1, 1049 Lisboa, Portugal}

\abstract{
We formulate a predictive model of fermion masses and mixings based on a $\Delta(27)$ family symmetry. In the quark sector the model leads to the viable mixing inspired texture where the Cabibbo angle comes from the down quark sector and the other angles come from both up and down quark sectors. In the lepton sector the model generates a predictive structure for charged leptons and, after radiative seesaw, an effective neutrino mass matrix with only one real and one complex parameter.
We carry out a detailed analysis of the predictions in the lepton sector, where the model is only viable for inverted neutrino mass hierarchy, predicting a strict correlation between $\theta_{23}$ and $\theta_{13}$. We show a benchmark point that leads to the best-fit values of $\theta_{12}$, $\theta_{13}$, predicting a specific $\sin^2\theta_{23} \simeq 0.51$ (within the $3 \sigma$ range), a leptonic CP-violating Dirac phase $\delta \simeq 281.6 ^\circ$ and for neutrinoless double-beta decay $m_{ee} \simeq 41.3$ meV.
We turn then to an analysis of the dark matter candidates in the model, which are stabilized by an unbroken $\mathbb{Z}_2$ symmetry. We discuss the possibility of scalar dark matter, which can generate the observed abundance through the Higgs portal by the standard WIMP mechanism. An interesting possibility arises if the lightest heavy Majorana neutrino is the lightest $\mathbb{Z}_2$-odd particle. The model can produce a viable fermionic dark matter candidate, but only as a feebly interacting massive particle (FIMP), with the smallness of the coupling to the visible sector protected by a symmetry and directly related to the smallness of the light neutrino masses.
}

\maketitle

\section{Introduction \label{intro}}

A well motivated extension of the Standard Model (SM) is adding a family symmetry in order to account for the observed pattern of SM
fermion masses and mixings, i.e. addressing the numerous
Yukawa couplings and the large hierarchy between them. These symmetries operate on the generations of fermions and tackle the flavour problem, one of the most relevant of the problems of the SM.
The details of the spontaneous breaking of the family symmetry lead to specific Yukawa structures and postdictions for the mixing angles in the quark or lepton sector. Recent reviews on discrete flavour
groups can be found in Refs.~\cite{Ishimori:2010au,Altarelli:2010gt,King:2013eh,King:2014nza,King:2017guk}. In particular the $\Delta(27)$ discrete group~
\cite{deMedeirosVarzielas:2006fc,Ma:2006ip, Ma:2007wu, Varzielas:2012nn,Bhattacharyya:2012pi,Ma:2013xqa,Nishi:2013jqa,Varzielas:2013sla,Aranda:2013gga,Varzielas:2013eta, Ma:2014eka, Abbas:2014ewa,Varzielas:2015aua,Abbas:2015zna, Chen:2015jta, Bjorkeroth:2015uou, Vien:2016tmh,Hernandez:2016eod,CarcamoHernandez:2017owh,deMedeirosVarzielas:2017sdv,CarcamoHernandez:2018iel}
has attracted a lot of attention as a promising family symmetry for explaining the observed pattern of SM fermion masses and mixing angles.

Another prominent issue in particle physics that motivates theories beyond the SM is its lack of a viable Dark Matter (DM) candidate.
In fact, there is compelling evidence for the existence of DM, an unknown, non-baryonic matter component whose abundance in the Universe exceeds the amount of ordinary matter roughly by a factor of five~\cite{Ade:2015xua}.
Still, the non-gravitational nature of DM remains a mystery~\cite{Bergstrom:2000pn, Bertone:2016nfn, deSwart:2017heh}.
Most prominent extensions of the SM feature Weakly Interacting Massive Particles (WIMPs) as DM.
WIMPs typically have order one couplings to the SM and masses at the electroweak scale.
The observation that this theoretical setup gives the observed relic abundance is the celebrated WIMP {\it miracle}~\cite{Arcadi:2017kky}.
In the standard WIMP paradigm, DM is a thermal relic produced by the freeze-out mechanism.
However, the observed DM abundance may have been generated also out of equilibrium by the so-called freeze-in mechanism~\cite{McDonald:2001vt, Choi:2005vq, Kusenko:2006rh, Hall:2009bx, Petraki:2007gq, Bernal:2017kxu}.
In this scenario, the DM particle couples to the visible SM sector very weakly, so that it never enters chemical equilibrium.
Due to the small coupling strength, the DM particles produced via the freeze-in mechanism have been called Feebly Interacting Massive Particles (FIMPs)~\cite{Hall:2009bx}; see Ref.~\cite{Bernal:2017kxu} for a recent review.

The solutions to the DM and the flavour problems have indeed often been approached separately in the literature.
Nevertheless one could entertain the idea that they have a common origin, 
whether because some residual flavour symmetry stabilizes it~\cite{Ma:2006km,Hernandez:2013dta,Campos:2014lla,Hernandez:2015hrt,Varzielas:2015joa,Arbelaez:2016mhg,Kownacki:2016hpm,Nomura:2016emz,Nomura:2016pgg,Nomura:2016dnf,Chulia:2016ngi,CarcamoHernandez:2016pdu,CarcamoHernandez:2017kra,CarcamoHernandez:2017cwi,Alvarado:2017bax}, or where there is a dark sector which communicates to the visible sector only through family symmetry mediators~\cite{Calibbi:2015sfa, Varzielas:2015sno}. 

With respect to the flavour problem, a viable form of the Yukawa structure for quarks is the mixing inspired texture where the Cabibbo angle originates from the down-quark sector and the remaining (smaller) mixing angles come from the more hierarchical up quark mixing~\cite{Hernandez:2014zsa}. We build a model based on the non-Abelian group $\Delta(27)$ which achieves a generalisation of this mixing inspired texture for the quarks, and is therefore phenomenologically viable. The model leads to a structure for the charged leptons which is diagonal apart from an entry mixing the first and third generations. The effective neutrino mass matrix arises through radiative seesaw and is in this case  a very simple structure, a sum of a democratic structure (all entries equal) plus a contribution only on the first diagonal entry. This predictive scenario for the leptons leads to a good fit to all masses and mixing angles with a correlation between $\theta_{13}$ and $\theta_{23}$, which depend only on the parameters of the charged lepton sector. In addition to the $\Delta(27)$, we need to employ $\mathbb{Z}_N$ symmetries that constrain the allowed terms, and within these, a single $\mathbb{Z}_2$ symmetry remains unbroken and stabilizes a DM, which can be either the lightest of the right-handed neutrinos (which are the only $\mathbb{Z}_2$-odd fermions) or a $\mathbb{Z}_2$-odd scalar. The model can lead to the correct relic abundance either under the WIMP or the FIMP scenarios.

\section{The Model \label{model}}

The model we 
propose is an extension of the SM that incorporates the $\Delta \left( 27\right) \times \mathbb{Z}_{2}\times \mathbb{Z}_{5}\times \mathbb{Z}_{6}\times \mathbb{Z}_{10}\times \mathbb{Z}_{16}$ discrete symmetry and a particle content extended with the SM singlets: scalars $\sigma$, $\eta_{1}$, $\eta_{2}$, $\rho$, $\Phi$, $\Xi$, $\varphi$ and two right handed Majorana neutrinos $N_{1,\,2R}$. All the non-SM fields are charged under the above mentioned discrete symmetry.
All the discrete groups are spontaneously broken, except for the $\mathbb{Z}_{2}$
under which only $\varphi$ and $N_{1,\,2R}$ are odd.
In this setup the light active neutrino masses arise 
at one-loop level through a radiative
seesaw mechanism, involving two right handed Majorana neutrinos and the $\mathbb{Z}_2$ odd scalars that do not acquire VEVs.

Our model reproduces a predictive mixing inspired textures 
where the Cabibbo mixing arises from the
down-type quark sector whereas the remaining mixing angles receive contributions from both up and down type quark sectors. These textures describe the charged fermion masses
and quark mixing pattern in terms of different powers of the Wolfenstein
parameter $\lambda =0.225$ and order one parameters. The full symmetry $%
\mathcal{G}$ of the model exhibits the following spontaneous breaking: 
\begin{eqnarray}
&&\mathcal{G}=SU(3)_{C}\times SU\left( 2\right) _{L}\times U\left( 1\right)
_{Y}\times \Delta \left( 27\right) \times \mathbb{Z}_{2}\times \mathbb{Z}_{5}\times \mathbb{Z}_{6}\times \mathbb{Z}_{10}\times \mathbb{Z}_{16}  \notag \\
&&\hspace{35mm}\Downarrow \Lambda  \notag \\[0.12in]
&&\hspace{15mm}SU(3)_{C}\times SU\left( 2\right) _{L}\times U\left( 1\right)
_{Y}\times \mathbb{Z}_{2}  \notag \\[0.12in]
&&\hspace{35mm}\Downarrow v  \notag \\[0.12in]
&&\hspace{15mm}SU(3)_{C}\times U\left( 1\right) _{Q}\times \mathbb{Z}_{2}\,,
\end{eqnarray}
where $\Lambda $ is the scale of breaking of the $\Delta \left( 27\right) \times \mathbb{Z}_{2}\times \mathbb{Z}_{5}\times \mathbb{Z}_{6}\times \mathbb{Z}_{10}\times \mathbb{Z}_{16}$ discrete group, which we assume
to be much larger than the electroweak symmetry breaking scale $v=246$~GeV.

The assignments of the scalars and the fermions under the 
$\Delta \left( 27\right) \times \mathbb{Z}_{2}\times \mathbb{Z}_{5}\times \mathbb{Z}_{6}\times \mathbb{Z}_{10}\times \mathbb{Z}_{16}$ discrete group
are listed in Tables~\ref{ta:scalars} and~\ref{ta:fermions}, where the dimensions of the $\Delta \left( 27\right) $ irreducible
representations are specified by numbers in boldface and different
charges are written in the additive notation. It is worth mentioning that all the 
scalar fields of the model acquire non-vanishing VEVs, except for the
SM singlet scalar field $\varphi$, which is the
only scalar charged under the preserved $\mathbb{Z}_{2}$ symmetry.

\begin{table}[tbp]
\begin{center}
\begin{tabular}{|c||c|c|c|c|c|c|c|c|}
\hline
& $\phi$ & $\sigma$ & $\eta_1$ & $\eta_2$ & $\rho$ & $\Phi$ & $\Xi$ & $\varphi$ \\ \hline\hline
$\Delta\left(27\right)$ & $\mathbf{1}_{\mathbf{0,0}}$ & $\mathbf{1}_{\mathbf{0,0}}$ & $\mathbf{1}_{\mathbf{0,0}}$ & $\mathbf{1}_{\mathbf{0,0}}$ & $\mathbf{1}_{\mathbf{0,1}}$ & $\overline{\mathbf{3}}$ & $\overline{\mathbf{3}}$ & $\mathbf{1}_{\mathbf{0,0}}$ \\ \hline
 $\mathbb{Z}_2$ & 0 & 0 & 0 & 0 & 0 & 0 & 0 & 1\\ \hline
 $\mathbb{Z}_5$ & 0 & 0 & 0 & 0 & 1 & 0 & 0 & 0\\ \hline
 $\mathbb{Z}_6$ & 0 & 0 & 0 & 0 & 0 & 1 & -2 & 0\\ \hline
 $\mathbb{Z}_{10}$ & 0 & 0 & -5 & -1 & 0 & 0 & 0 & 0\\ \hline
 $\mathbb{Z}_{16}$ & 0 & -1 & 0 & -1 & 0 & 0 & 0 & 0\\
\hline
\end{tabular}
\caption{Scalar assignments under $\Delta \left( 27\right) \times \mathbb{Z}_{2}\times \mathbb{Z}_{5}\times \mathbb{Z}_{6}\times \mathbb{Z}_{10}\times \mathbb{Z}_{16}$. The scalar $\phi$ corresponds to the SM $SU(2)$ Higgs doublet. The $\mathbb{Z}_{N}$ charges, $q$, shown in the additive notation so that the group element is $\omega = e^{2 \pi i\, q/N }$. For  the $\Delta(27)$ representations and the notations see Appendix~\ref{A}.}
\label{ta:scalars}
\end{center}
\end{table}
\begin{table}[tbp]
\resizebox{15cm}{!}{
\renewcommand{\arraystretch}{1.2}
\begin{tabular}{|c||c|c|c|c|c|c|c|c|c|c|c|c|c|c|c|}
\hline
& $q_{1L}$ & $q_{2L}$ & $q_{3L}$ & $u_{1R}$ & $u_{2R}$ & $u_{3R}$ & $d_{1R}$
& $d_{2R}$ & $d_{3R}$ & $l_{L}$ & $l_{1R}$ & $l_{2R}$ & $l_{3R}$ & $N_{1R}$
& $N_{2R}$ \\ \hline\hline
$\Delta\left(27\right)$ & $\mathbf{1}_{\mathbf{0,0}}$ & $\mathbf{1}_{\mathbf{%
0,0}}$ & $\mathbf{1}_{\mathbf{0,0}}$ & $\mathbf{1}_{\mathbf{0,0}}$ & $%
\mathbf{1}_{\mathbf{0,0}}$ & $\mathbf{1}_{\mathbf{0,0}}$ & $\mathbf{1}_{%
\mathbf{0,0}}$ & $\mathbf{1}_{\mathbf{0,0}}$ & $\mathbf{1}_{\mathbf{0,0}}$ & 
$\overline{\mathbf{3}}$ & $\mathbf{1}_{\mathbf{0,0}}$ & $\mathbf{1}_{\mathbf{0,0}}$ & $\mathbf{1}_{\mathbf{0,0}}$ & $\mathbf{1}_{\mathbf{0,0}}$ & $%
\mathbf{1}_{\mathbf{0,0}}$ \\ \hline
$\mathbb{Z}_2$ & 0 & 0 & 0 & 0 & 0 & 0 & 0 & 0 & 0 & 0 & 0 & 0 & 0 & 1 & 1
\\ \hline
$\mathbb{Z}_5$ & 0 & 0 & 0 & 0 & 0 & 0 & 0 & 0 & 0 & 0 & 0 & -1 & 3 & 0 & 0
\\ \hline
$\mathbb{Z}_{6}$ & 0 & 0 & 0 & 0 & 0 & 0 & 0 & 0 & 0 & 1 & 0 & 0 & 0 & 0
& 3 \\ \hline
$\mathbb{Z}_{10}$ & 0 & 0 & 0 & 0 & 0 & 0 & 0 & 5 & 0 & 0 & 0 & 0 & 0 & 0
& 0 \\ \hline
$\mathbb{Z}_{16}$ & -4 & -2 & 0 & 4 & 2 & 0 & 3 & 2 & 3 & 0 & 8 & 3 & 0 & 0
& 0 \\ \hline
\end{tabular}}%
\caption{The same as in Table~\ref{ta:scalars} but for fermions.}
\label{ta:fermions}
\end{table}

With the above particle content, the following quark, charged lepton and
neutrino Yukawa terms arise: 

\begin{eqnarray}
\tciLaplace _{Y}^{\left( U\right) } &=&y_{11}^{\left( U\right) }\overline{q}%
_{1L}\widetilde{\phi }u_{1R}\frac{\sigma ^{8}}{\Lambda ^{8}}+y_{12}^{\left(
U\right) }\overline{q}_{1L}\widetilde{\phi }u_{2R}\frac{\sigma ^{6}}{\Lambda
^{6}}+y_{13}^{\left( U\right) }\overline{q}_{1L}\widetilde{\phi }u_{3R}\frac{%
\sigma ^{4}}{\Lambda ^{4}}  \notag \\
&&+y_{21}^{\left( U\right) }\overline{q}_{2L}\widetilde{\phi }u_{1R}\frac{%
\sigma ^{6}}{\Lambda ^{6}}+y_{22}^{\left( U\right) }\overline{q}_{2L}%
\widetilde{\phi }u_{2R}\frac{\sigma ^{4}}{\Lambda ^{4}}+y_{23}^{\left(
U\right) }\overline{q}_{2L}\widetilde{\phi }u_{3R}\frac{\sigma ^{2}}{\Lambda
^{2}}  \notag \\
&&+y_{31}^{\left( U\right) }\overline{q}_{3L}\widetilde{\phi }u_{1R}\frac{%
\sigma ^{4}}{\Lambda ^{4}}+y_{32}^{\left( U\right) }\overline{q}_{3L}%
\widetilde{\phi }u_{2R}\frac{\sigma ^{2}}{\Lambda ^{2}}+y_{33}^{\left(
U\right) }\overline{q}_{3L}\widetilde{\phi }u_{3R}+h.c,
\label{lyu}
\end{eqnarray}
\begin{eqnarray}
\tciLaplace _{Y}^{\left( D\right) } &=&y_{11}^{\left( D\right) }\overline{q}%
_{1L}\phi d_{1R}\frac{\sigma ^{7}}{\Lambda ^{7}}+y_{12}^{\left( D\right) }%
\overline{q}_{1L}\phi d_{2R}\frac{\eta _{2}^{5}\sigma }{\Lambda ^{6}}%
+y_{13}^{\left( D\right) }\overline{q}_{1L}\phi d_{3R}\frac{\sigma ^{7}}{%
\Lambda ^{7}}  \notag \\
&&+y_{21}^{\left( D\right) }\overline{q}_{2L}\phi d_{1R}\frac{\sigma ^{5}}{%
\Lambda ^{5}}+y_{22}^{\left( D\right) }\overline{q}_{2L}\phi d_{2R}\frac{%
\sigma ^{4}\eta _{1}}{\Lambda ^{5}}+y_{23}^{\left( D\right) }\overline{q}%
_{2L}\phi d_{3R}\frac{\sigma ^{5}}{\Lambda ^{5}}  \notag \\
&&+y_{31}^{\left( D\right) }\overline{q}_{3L}\phi d_{1R}\frac{\sigma ^{3}}{%
\Lambda ^{3}}++y_{32}^{\left( D\right) }\overline{q}_{3L}\phi d_{2R}\frac{%
\eta _{1}\sigma ^{2}}{\Lambda ^{3}}+y_{33}^{\left( D\right) }\overline{q}%
_{3L}\phi d_{3R}\frac{\sigma ^{3}}{\Lambda ^{3}}+h.c,
\label{lyd}
\end{eqnarray}
\begin{eqnarray}
\tciLaplace _{Y}^{\left( l\right) }&=&y_{33}^{\left( l\right) }\left( 
\overline{l}_{L}\phi \Phi \right) _{\mathbf{1}_{\mathbf{0,1}}}l_{3R}\frac{%
\rho ^{2}}{\Lambda ^{3}}+y_{13}^{\left( l\right) }\left( \overline{l}%
_{L}\phi \Phi \right) _{\mathbf{1}_{\mathbf{0,0}}}l_{3R}\frac{\left( \rho
^{\ast }\right) ^{3}}{\Lambda ^{4}}\notag\\
&&+y_{22}^{\left( l\right) }\left(\overline{l}_{L}\phi \Phi \right) _{\mathbf{1}_{\mathbf{0,2}}}l_{2R}\frac{%
\rho \sigma ^{3}}{\Lambda ^{5}}+y_{11}^{\left( l\right) }\left( \overline{l}%
_{L}\phi \Phi \right) _{\mathbf{1}_{\mathbf{0,0}}}l_{1R}\frac{\sigma ^{8}}{%
\Lambda ^{9}}+h.c,
\label{lyl}
\end{eqnarray}
\begin{eqnarray}
\tciLaplace _{Y}^{\left( \nu \right) }&=&y_{1}^{\left( \nu \right) }\left( 
\overline{l}_{L}\widetilde{\phi }\Phi \right) _{\mathbf{1}_{\mathbf{0,0}%
}}N_{1R}\frac{\varphi }{\Lambda ^{2}}+y_{2}^{\left( \nu \right) }\left( 
\overline{l}_{L}\widetilde{\phi }\Xi \right) _{\mathbf{1}_{\mathbf{0,0}%
}}N_{2R}\frac{\varphi }{\Lambda ^{2}}\nonumber\\
&&\quad\quad+m_{N_{1R}}\overline{N}%
_{1R}N_{1R}^{C}+m_{N_{2R}}\overline{N}_{2R}N_{2R}^{C}+h.c,  \label{Lyn}
\end{eqnarray}
where the dimensionless couplings in Eqs.~(\ref{lyu})-(\ref{Lyn}) are $%
\mathcal{O}(1)$ parameters, which we will constrain through a fit to the
observed fermion masses and mixings parameters.

In addition to these terms, the symmetries unavoidably allow terms in $\tciLaplace _{Y}^{\left( l\right) }$ where the contraction
$\left( \overline{l}_{L}\phi \Phi \right)$ is replaced with
$\left( \overline{l}_{L}\phi \Xi \right) \left( \Xi^\dagger \Phi \right)$. For example, in addition to 
$\left( \overline{l}_{L}\phi \Phi \right) l_{3R}\frac{%
\rho ^{2}}{\Lambda ^{3}}$, the following term is allowed:
$\left( \overline{l}_{L}\phi \Xi \right) \left( \Xi^\dagger \Phi \right) l_{3R}\frac{%
\rho ^{2}}{\Lambda ^{5}}$. These terms have two additional suppressions of $\langle \Xi \rangle/\Lambda$ and can be safely neglected if there is a mild hierarchy between $\langle \Xi \rangle$ and $\langle \Phi \rangle$. This hierarchy in the VEVs is consistent is also consistent with the mild hierarchy obtained for the masses of the light effective neutrinos after seesaw.

As indicated by the current low energy quark flavour data encoded in the
Standard parametrization of the quark mixing matrix, the complex phase
responsible for CP violation in the quark sector is associated with the
quark mixing angle in the $1$-$3$ plane. Consequently, in order to reproduce
the experimental values of quark mixing angles and CP violating phase, the Yukawa coupling in Eq.~(\ref{lyu}) $y_{13}^{\left( U\right) }$ is required to be complex.

An explanation of the role of each discrete group factor of our model is
provided in the following. The $\Delta \left( 27\right)$, $\mathbb{Z}_{5}$, $\mathbb{Z}_{6}$
and $\mathbb{Z}_{10}$ discrete groups are crucial for reducing the number of
model parameters, thus increasing the predictivity of our model and giving
rise to predictive and viable textures for the fermion sector, consistent
with the observed pattern of fermion masses and mixings, as will be shown
later in Sects.~\ref{quarksector} and~\ref{leptonsector}. The $\Delta \left(
27\right) $, $\mathbb{Z}_{5}$, $\mathbb{Z}_{6}$ and $\mathbb{Z}_{10}$ discrete groups, which
are spontaneously broken, determine the allowed entries of the quark mass
matrices as well as their hierarchical structure in terms of different
powers of the Wolfenstein parameter, thus giving rise to the observed SM
fermion mass and mixing pattern. In particular the $\mathbb{Z}_{5}$ discrete symmetry is crucial for explaining the tau and muon charged lepton masses as well as the Cabbibo sized value for the reactor mixing angle, which only arises from the charged lepton sector. The $\mathbb{Z}_{6}$ discrete group allows us to get a predictive texture for the light active neutrino sector. This symmetry forbids mixings between the two right handed Majorana neutrinos $N_{1R}$ and $N_{2R}$. The $\mathbb{Z}_{10}$ discrete symmetry allows to get the right hierarchical in the second column of the down type quark mass matrix crucial to successfully reproduce the right values of the strange quark mass and the Cabbibo angle with $\mathcal{O}(1)$ parameters.

As a result of the $\Delta \left( 27\right) \times \mathbb{Z}_{2}\times \mathbb{Z}_{5}\times \mathbb{Z}_{6}\times \mathbb{Z}_{10}\times \mathbb{Z}_{16}$ charge assignment for scalars and quarks given in Tables~\ref{ta:scalars} and~\ref{ta:fermions}, the Cabibbo mixing will arise from the down type quark sector, whereas the remaining mixing angles will receive contributions for both up and down type sectors. The preserved $\mathbb{Z}_{2}$ symmetry allows the implementation of
the one loop level radiative seesaw mechanism for the generation of the
light active neutrino masses as well as provides a viable DM particle candidate.

We assume the following VEV pattern for the $\Delta \left( 27\right) $
triplet SM singlet scalars 
\begin{equation}
\label{eq:Phi-Sigma-VEVs}
\left\langle \Phi \right\rangle =v_{\Phi }\,\left( 1,0,0\right)\, ,\hspace{1.5cm}%
\left\langle \Xi \right\rangle =v_{\Xi }\,\left( 1,1,1\right)\, ,
\end{equation}
which is consistent with the scalar potential minimization equations for a
large region of parameter space as shown in detail in Ref.~\cite{deMedeirosVarzielas:2017glw}.

Besides that, as the hierarchy among charged fermion masses and quark mixing
angles emerges from the breaking of the $\Delta \left( 27\right) \times \mathbb{Z}_{2}\times \mathbb{Z}_{5}\times \mathbb{Z}_{6}\times \mathbb{Z}_{10}\times \mathbb{Z}_{16}$ discrete group, we set the VEVs of the
SM singlet scalar fields with respect to the Wolfenstein parameter $\lambda
=0.225$ and the model cutoff $\Lambda $, as follows:
\begin{equation}
v_{\sigma}\sim v_{\eta_{1}}\sim v_{\eta _{2}}\sim v_{\rho}\sim v_{\Phi }\sim \lambda\,\Lambda\,, v_{\Xi } \sim \lambda^{3/2}\,\Lambda\,.  \label{VEVs}
\end{equation}
We require a mild hierarchy between the VEVs of the two $\Delta(27)$ triplet scalars $\Phi$ and $\Xi$ (merely a factor of two), which is sufficient to suppress the effect of unavoidable terms in the charged lepton sector, which could otherwise spoil the phenomenology of the model discussed in Section \ref{leptonsector}.
The model cutoff scale $\Lambda $ can be thought of as the scale of the UV
completion of the model, e.g. the masses of Froggatt-Nielsen messenger
fields. It is straightforward to show that the assumption regarding the VEV size of the SM singlet scalars given by Eq.~\eqref{VEVs} is consistent with the scalar potential minimization. That assumption given by that equation can be justified by considering $\mu^2_{\Xi}<\mu^2_{\sigma}\sim\mu^2_{\eta_{1}}\sim\mu^2_{\eta_{2}}\sim\mu^2_{\rho}\sim\mu^2_{\Phi}$ and the quartic scalar couplings of the same order of magnitude.

\section{Quark Masses and Mixings \label{quarksector}}

From the quark Yukawa terms of Eqs.~(\ref{lyu}) and~(\ref{lyd}), inserting the
VEV magnitudes of the scalars with respect to $\Lambda$ we rewrite it in
term of effective parameters
\begin{eqnarray}
\tciLaplace _{Y}^{\left( Q\right) } &=&\frac{(v+H)}{\sqrt 2} \left(
a_{11}^{\left( U\right) }\overline{q}_{1L}u_{1R}\lambda^{8}+a_{12}^{\left( U\right) }%
\overline{q}_{1L}u_{2R}\lambda^6+a_{13}^{\left( U\right) }\overline{q}_{1L}
u_{3R}\lambda^{4}\nonumber\right.\\
&&+\left.a_{21}^{\left( U\right) }\overline{q}_{2L}u_{1R}\lambda^{6}+a_{22}^{\left( U\right) }\overline{q}_{2L}u_{2R}\lambda^{4}+a_{23}^{\left( U\right) }\overline{q}_{2L}u_{3R}\lambda^{2}\nonumber\right.\\
&&+\left. a_{31}^{\left( U\right) }\overline{q}_{3L}u_{1R}\lambda^{4}+a_{32}^{\left( U\right) }%
\overline{q}_{2L}u_{3R} \lambda^2+a_{33}^{\left( U\right) }\overline{q}_{3L}u_{3R}\right)\nonumber\\
&&+\frac{(v+H)}{\sqrt 2} \left(
a_{11}^{\left( D\right) }\overline{q}_{1L}d_{1R}\lambda^{7}+a_{12}^{\left( D\right) }%
\overline{q}_{1L}d_{2R}\lambda^6+a_{13}^{\left( U\right) }\overline{q}_{1L}
d_{3R}\lambda^{7}\nonumber\right.\\
&&+\left.a_{21}^{\left(D\right) }\overline{q}_{2L}d_{1R}\lambda^{5}+a_{22}^{\left(D\right) }\overline{q}_{2L}d_{2R}\lambda^{5}+a_{23}^{\left( D\right) }\overline{q}_{2L}d_{3R}\lambda^{5}\nonumber\right.\\
&&+\left. a_{31}^{\left( D\right) }\overline{q}_{3L}d_{1R}\lambda^3+a_{32}^{\left( D\right) }%
\overline{q}_{3L}d_{2R}\lambda^3+a_{33}^{\left(D\right) }\overline{q}_{3L}d_{3R}\lambda^3\right)+h.c.
\end{eqnarray}
Then it follows that the quark mass matrices take the form: 
\begin{equation}
M_{U}=\left( 
\begin{array}{ccc}
a_{11}^{\left( U\right) }\lambda ^{8} & a_{12}^{\left( U\right) }\lambda ^{6} & a_{13}^{\left( U\right) }\lambda
^{4} \\ 
a_{21}^{\left( U\right) }\lambda ^{6} & a_{2}^{\left( U\right) }\lambda ^{4} & a_{23}^{\left( U\right) }\lambda
^{2} \\ 
a_{31}^{\left( U\right) }\lambda^{4} & a_{32}^{\left( U\right) }\lambda^{2} & a_{33}^{\left( U\right) }%
\end{array}%
\right) \frac{v}{\sqrt{2}},\hspace{1.0cm}M_{D}=\left( 
\begin{array}{ccc}
a_{11}^{\left( D\right) }\lambda ^{7} & a_{12}^{\left( D\right) }\lambda ^{6}
& a_{13}^{\left( D\right) }\lambda ^{7} \\ 
a_{21}^{\left( D\right) }\lambda ^{5} & a_{22}^{\left( D\right) }\lambda ^{5} & a_{23}^{\left( D\right) }\lambda ^{5} \\ 
a_{31}^{\left( D\right) }\lambda ^{3} & a_{32}^{\left( D\right) }\lambda ^{3} & a_{33}^{\left( D\right) }\lambda ^{3}
\end{array}%
\right) \frac{v}{\sqrt{2}},
\label{Quarktextures}
\end{equation}
where $a_{ij}^{\left( U\right) }$ and $a_{ij}^{\left(
D\right) }$ ($i,\,j=1,\,2,\,3$) are $\mathcal{O}(1)$ parameters. Here $%
\lambda =0.225$ is the Wolfenstein parameter and $v=246$ GeV the
scale of electroweak symmetry breaking. The SM quark mass textures given
above indicate that the Cabibbo mixing emerges from the down type quark
sector, whereas the remaining mixing angles receive contributions from both up and down type quark sectors. Indeed, this texture is a generalisation of the particular case referred to as the mixing inspired texture~\cite{Hernandez:2014zsa}, in which the two small quark mixing angles would arise solely from the up type quark sector. Besides that, the low energy quark flavour data indicates that the CP violating phase in the quark sector is associated with the quark mixing
angle in the 1-3 plane, as follows from the Standard parametrization of the
quark mixing matrix.  Consequently, in order to get quark mixing angles and a
CP violating phase consistent with the experimental data, we 
adopt a minimalistic scenario where all the dimensionless parameters given in Eq.~(\ref{Quarktextures}) are real, except for $a_{13}^{\left( U\right)}$, taken to
be complex.

The obtained values for the physical quark mass spectrum~\cite%
{Bora:2012tx,Xing:2007fb}, mixing angles and Jarlskog invariant~\cite%
{Olive:2016xmw} are consistent with their experimental data, as shown in
Table~\ref{Tab}, starting from the following benchmark point that would correspond to the limit of the mixing inspired texture \cite{Hernandez:2014zsa} \footnote{This limit corresponds to $a_{12}^{\left( U\right) } = a_{21}^{\left( U\right) } =a_{31}^{\left( U\right) } =a_{32}^{\left( U\right) } =0$, $a_{13}^{\left( D\right) }=a_{21}^{\left( D\right) } =a_{23}^{\left( D\right) }=a_{31}^{\left( D\right) }=a_{32}^{\left( D\right) }=0$.
  }:
\begin{eqnarray}
a_{11}^{\left( U\right) } &\simeq &1.266,\hspace{0.4cm}a_{22}^{\left( U\right)}\simeq 1.430,\hspace{0.4cm}a_{33}^{\left(U\right) }\simeq 0.989,\hspace{0.4cm}a_{13}^{\left( U\right) }\simeq -0.510-1.262i,\hspace{0.4cm}a_{23}^{\left( U\right) }\simeq 0.806,  \notag \\
a_{11}^{\left( D\right) } &\simeq &0.550,\hspace{0.4cm}a_{22}^{\left( D\right)
}\simeq 0.554,\hspace{0.4cm}a_{33}^{\left( D\right) }\simeq 1.411,\hspace{0.4cm}%
a_{12}^{\left( D\right) }\simeq 0.565.
\label{fit-q}
\end{eqnarray}
\begin{table}[tbh]
\begin{center}
\begin{tabular}{|c||l|l|}
\hline
Observable & Model value & Experimental value \\ \hline\hline
$m_{u}$ [MeV]& \quad $1.47$ & \quad $1.45_{-0.45}^{+0.56}$ \\ \hline
$m_{c}$ [MeV]& \quad $641$ & \quad $635\pm 86$ \\ \hline
$m_{t}$ [GeV]& \quad $172$ & \quad $172.1\pm 0.6\pm 0.9$ \\ \hline
$m_{d}$ [MeV]& \quad $2.8$ & \quad $2.9_{-0.4}^{+0.5}$ \\ \hline
$m_{s}$ [MeV]& \quad $57.5$ & \quad $57.7_{-15.7}^{+16.8}$ \\ \hline
$m_{b}$ [GeV]& \quad $2.81$ & \quad $2.82_{-0.04}^{+0.09}$ \\ \hline
$\sin\theta^{(q)}_{12}$ & \quad $0.225$ & \quad $0.225$ \\ \hline
$\sin\theta^{(q)}_{23}$ & \quad $0.0414$ & \quad $0.0414$ \\ \hline
$\sin\theta^{(q)}_{13}$ & \quad $0.00355$ & \quad $0.00355$ \\ \hline
$\delta $ & \quad $68^{\circ }$ & \quad $68^{\circ }$ \\ \hline
\end{tabular}%
\end{center}
\caption{Model and experimental values of the quark masses and CKM
parameters.}
\label{Tab}
\end{table}
In Table~\ref{Tab} we show the model and experimental values for the
physical observables of the quark sector. We use the $M_{Z}$-scale
experimental values of the quark masses given by Ref.~\cite{Bora:2012tx}
(which are similar to those in Ref.~\cite{Xing:2007fb}). The experimental values
of the CKM parameters are taken from Ref.~\cite{Agashe:2014kda}. As
indicated by Table~\ref{Tab}, the obtained quark masses, quark mixing
angles, and CP violating phase can be fitted to the
experimental low energy quark flavour data. 
We note that the values (\ref{fit-q}) of the parameters $a^{(U,\,D)}_{i}$ are compatible 
with $\mathcal{O}(1)$. This fact supports the desired feature of the model that the hierarchy of masses and mixing angles are encoded in the powers of $\lambda$ and texture zero of the mass matrices Eq.~(\ref{Quarktextures}), which in its turn is the consequence of the particular flavour symmetry of the model.

\section{Lepton Masses and Mixings \label{leptonsector}}

We can expand the contractions of the $\Delta (27)$ (anti-)triplets $l_{L}$, 
$\Phi $ and $\Xi $
according to the scalar VEV directions in Eq.~(\ref{eq:Phi-Sigma-VEVs}).
Then we 
have $\left( \overline{l}_{L}\Phi
\right) _{\mathbf{1}_{\mathbf{0,0}}}\propto \overline{l}_{1L}$, $\left( 
\overline{l}_{L}\Phi \right) _{\mathbf{1}_{\mathbf{0,2}}}\propto \overline{l}%
_{2L}$, $\left( \overline{l}_{L}\Phi \right) _{\mathbf{1}_{\mathbf{0,1}%
}}\propto \overline{l}_{3L}$, and $\left( \overline{l}_{L}\Xi \right) _{%
\mathbf{1}_{\mathbf{0,0}}}\propto \left( \overline{l}_{1L}+\overline{l}_{2L}+%
\overline{l}_{3L}\right) $. 
Taking into account $v_{\Phi}\sim v_{\Sigma} \sim \lambda\,\Lambda$, specified in Eq.~(\ref{VEVs}), we rewrite Eqs.~(\ref{lyl}) and~(\ref{Lyn}) in the form
\begin{equation}
\label{eq:chlept-1}
\tciLaplace _{Y}^{\left( l\right) }=\frac{(v+H)}{\sqrt{2}}\left(
a_{3}^{\left( l\right) }\overline{l}_{3L}l_{3R}\lambda ^{3}+a_{4}^{\left(
l\right) }\overline{l}_{1L}l_{3R}\lambda ^{4}+a_{2}^{\left( l\right) }%
\overline{l}_{2L}l_{2R}\lambda ^{5}+a_{1}^{\left( l\right) }\overline{l}%
_{1L}l_{1R}\lambda ^{9} \right) +h.c ,
\end{equation}%
\begin{eqnarray}
\tciLaplace _{Y}^{\left( \nu \right) }&=&\frac{(v+H)}{\sqrt{2}} \left( y_{2}^{\left( \nu \right) }v_{\Xi
}\left( \overline{l}_{1L}+\overline{l}_{2L}+\overline{l}_{3L}\right) 
N_{2R}\frac{\varphi }{\Lambda ^{2}}+y_{1}^{\left( \nu
\right) }v_{\Phi }\overline{l}_{1L} N_{1R}\frac{\varphi }{%
\Lambda ^{2}} \right) \nonumber \\
&&\qquad+m_{N_{1R}}\overline{N}_{1R}N_{1R}^{C}+m_{N_{2R}}\overline{N}%
_{2R}N_{2R}^{C} +h.c.
\label{eq:expanded_neutrinoL}
\end{eqnarray}
From Eq.~(\ref{eq:chlept-1}) we find the charged lepton mass matrix
\begin{equation}
M_{l}=\left( 
\begin{array}{ccc}
a_{1}^{\left( l\right) }\lambda ^{9} & 0 & a_{4}^{\left( l\right) }\lambda
^{4} \\ 
0 & a_{2}^{\left( l\right) }\lambda ^{5} & 0 \\ 
0 & 0 & a_{3}^{\left( l\right) }\lambda ^{3}%
\end{array}%
\right) \frac{v}{\sqrt{2}},
\end{equation}%
where $a_{k}^{\left( l\right) }$ ($k=1,\cdots ,4$) are $\mathcal{O}(1)$
dimensionless parameters.
The contribution from the charged lepton sector to the PMNS matrix, $U^{\left( l\right) }$ consists in a rotation by a single non-vanishing angle $\theta_{13}^{\left( l\right) }$ which depends crucially on $a_{4}^{\left( l\right)}$.

\begin{figure}[tbh]
\begin{center}
\vspace{-0.8cm} \resizebox{15cm}{21cm}
{\includegraphics[scale=0.5]{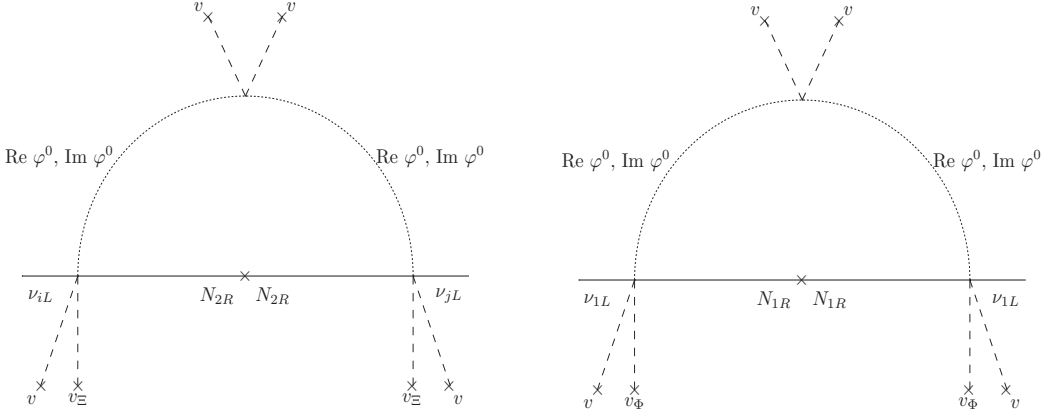}}{} 
\vspace{-14cm}
\caption{Loop Feynman diagrams contributing to the entries of the neutrino
mass matrix. Here $i,j=1,2,3$. The cross marks in the internal lines with $%
N_{nR}-N_{nR}$ ($n=1,2$) denote the insertion of the mass $m_{N_{nR}}$.}
\label{Loopdiagram}
\end{center}
\end{figure}

The effective neutrino mass matrix $M_\nu$ arises after radiative seesaw,
from the Yukawa terms (which we expanded in Eq.~(\ref{eq:expanded_neutrinoL})) with scalar $\varphi $ (which does not acquire a VEV)
and the masses of the right-handed neutrinos. The mechanism is associated with the loop diagrams in Fig.~\ref{Loopdiagram}.
Considering these diagrams and the Dirac couplings in Eq.(\ref{eq:expanded_neutrinoL}) with $\varphi$, which we represent in the matrix form $Y^\nu_\varphi$:
\begin{equation}
Y^{\nu }_{\varphi} = \frac{v}{\sqrt{2} \Lambda^2} \left( 
\begin{array}{cc}
v_\Phi y^{(\nu)}_1 & v_\Xi y^{(\nu)}_2 \\ 
0 & v_\Xi y^{(\nu)}_2 \\ 
0 & v_\Xi y^{(\nu)}_2 %
\end{array}%
\right) ,
\end{equation}
one reads off there will be a democratic contribution associated with the $y^{(\nu)}_2$ coupling filling each entry in $M_\nu$ equally (due to the coupling to the combination $\left( \overline{l}_{1L}+\overline{l}_{2L}+\overline{l}_{3L}\right) $) whereas the $y^{(\nu)}_1$ coupling is responsible for a contribution solely to the $11$ entry of $M_\nu$. 
Thus we write the effective neutrino mass matrix in the form
\begin{equation}
\label{eq:Mnu-1}
M_{\nu }=\left( 
\begin{array}{ccc}
A_{1}^{\left( \nu \right) } & A_{2}^{\left( \nu \right) } & A_{2}^{\left(
\nu \right) } \\ 
A_{2}^{\left( \nu \right) } & A_{2}^{\left( \nu \right) } & A_{2}^{\left(
\nu \right) } \\ 
A_{2}^{\left( \nu \right) } & A_{2}^{\left( \nu \right) } & A_{2}^{\left(
\nu \right) }%
\end{array}%
\right) ,
\end{equation}
where the dimensionful parameters $A_{1}^{\left( \nu \right) }$ and $%
A_{2}^{\left( \nu \right) }$ follow from the loop functions 
of the diagrams in Fig.~\ref{Loopdiagram}.
\begin{eqnarray}
A_{1}^{\left( \nu \right) }&\simeq& \frac{\left( y_{2}^{\left( \nu \right)
}\right) ^{2}v_{\Xi }^{2}v^{2}m_{N_{2R}}}{32\pi ^{2}\Lambda ^{4}}f\left( m_{%
\func{Re}\varphi },m_{\func{Im}\varphi },m_{N_{2R}}\right)\nonumber\\
&&\qquad+\frac{\left(
y_{1}^{\left( \nu \right) }\right) ^{2}v_{\Phi }^{2}v^{2}m_{N_{1R}}}{32\pi
^{2}\Lambda ^{4}}f\left( m_{\func{Re}\varphi },m_{\func{Im}\varphi
},m_{N_{1R}}\right) ,  \label{A1nu}
\end{eqnarray}
\begin{equation}
A_{2}^{\left( \nu \right) }\simeq \frac{\left( y_{2}^{\left( \nu \right)
}\right) ^{2}v_{\Xi }^{2}v^{2}m_{N_{2R}}}{32\pi ^{2}\Lambda ^{4}}f\left( m_{%
\func{Re}\varphi },m_{\func{Im}\varphi },m_{N_{2R}}\right) ,  \label{A2nu}
\end{equation}
\begin{equation}
f\left( m_{\func{Re}\varphi },m_{\func{Im}\varphi },m_{N_{kR}}\right) =\frac{%
m_{\func{Re}\varphi }^{2}}{m_{\func{Re}\varphi }^{2}-m_{N_{kR}}^{2}}\ln
\left( \frac{m_{\func{Re}\varphi }^{2}}{m_{N_{kR}}^{2}}\right) -\frac{m_{%
\func{Im}\varphi }^{2}}{m_{\func{Im}\varphi }^{2}-m_{N_{kR}}^{2}}\ln \left( 
\frac{m_{\func{Im}\varphi }^{2}}{m_{N_{kR}}^{2}}\right)\,,
\label{f}
\end{equation}
with $k=1$, $2$.
We note that $\varphi $ needs to be a complex scalar otherwise the loop
functions vanish, and further the real and imaginary parts of $\varphi $
must not have degenerate masses.
The structure of $M_{\nu}$ is such that it has an eigenvector  $(0,1,-1)/\sqrt{2}$ with a vanishing eigenvalue, corresponding therefore to a massless neutrino. This means the neutrino sector's contribution to the PMNS matrix, $U^{\left( \nu \right) }$, has one direction which is $(0,1,-1)/\sqrt{2}$, meaning $\theta_{13}^{\left( \nu \right) } =0 $ and $\theta_{23}^{\left(  \nu \right) }= \pi/4$. This gets modified by the contribution from the charged lepton sector such that the reactor angle is non-zero, but given that the associated state is the massless state this structure is viable for the inverted hierarchy of neutrino masses (but not for the normal hierarchy).
Indeed, we find that for our model the normal
hierarchy scenario leads to a too large reactor mixing angle, thus being ruled out by the
current data on neutrino oscillation experiments.

The $\mathcal{O}(1)$ dimensionless couplings $a_{i}^{\left( l\right) }$ ($%
i=1,\cdots ,4$) determine the charged lepton masses, the reactor mixing
parameter $\sin ^{2}\theta _{13} \neq 0$ and the deviation $\sin ^{2}\theta _{23} - 1/2 \neq 0$, which are correlated:
\begin{equation}
\sin ^{2}\theta _{23} = \frac{1}{2\,(1-\sin ^{2}\theta _{13})}\,.
\end{equation}
In turn, $A_{1}^{\left( \nu \right) }$
and $A_{2}^{\left( \nu \right) }$ are dimensionful  parameters crucial to
determine the neutrino mass squared splittings as well as the solar angle $\sin
^{2}\theta _{12}$.
For the sake of simplicity and proving these leptonic structures are viable, we assume that the parameters $a_{l}^{\left(
l\right) }$ ($l=1,\cdots ,4$), $A_{2}^{\left( \nu \right) }$ are real
whereas $A_{1}^{\left( \nu \right) }$ is taken to be complex.
We have
checked numerically that the simplest scenario of all lepton parameters ($%
a_{l}^{\left( l\right) }$ ($l=1,\cdots ,4$), $A_{1}^{\left( \nu \right) }$%
and $A_{2}^{\left( \nu \right) }$) being real leads to a solar mixing parameter $%
\sin ^{2}\theta _{12}$ close to about $0.2$, which is below its $3\sigma$ experimental lower bound.

In order to reproduce the experimental values of the physical observables of
the lepton sector, i.e. the three charged lepton masses, two neutrino mass
squared splittings and the three leptonic mixing parameters, we proceed to
fit the parameters $a_{k}^{\left( l\right) }$ ($k=1,\cdots ,4$), $%
\left\vert A_{1}^{\left( \nu \right) }\right\vert $,$\ A_{2}^{\left( \nu
\right) }$ and $\arg \left[ A_{1}^{\left( \nu \right) }\right] $. 

For the case of inverted neutrino mass hierarchy we find
the following best fit result 
\begin{eqnarray}
a_{1}^{\left( l\right) } &\simeq &1.936,\hspace{1cm}a_{2}^{\left( l\right)
}\simeq 1.025,\hspace{1cm}a_{3}^{\left( l\right) }\simeq 0.864,\hspace{1cm}%
a_{4}^{\left( l\right) }\simeq 0.813,  \notag \\
\left\vert A_{1}^{\left( \nu \right) }\right\vert &\simeq &69.7\hspace{0.5mm}%
\mbox{meV},\hspace{1cm}A_{2}^{\left( \nu \right) }\simeq 20.6\hspace{0.5mm}%
\mbox{meV},\hspace{1cm}\arg \left[ A_{1}^{\left( \nu \right) }\right] \simeq
-58.26^{\circ }.  \label{Fitpointneutrino}
\end{eqnarray}
The small hierarchy between effective parameters $\left\vert A_{1}^{\left( \nu \right) }\right\vert $,$\ A_{2}^{\left( \nu
\right) }$ is consistent with the mild hierarchy between $\langle \Phi \rangle$ and $\langle \Xi \rangle$.

\begin{figure}[t!]
\centering
\includegraphics[width=0.47\textwidth]{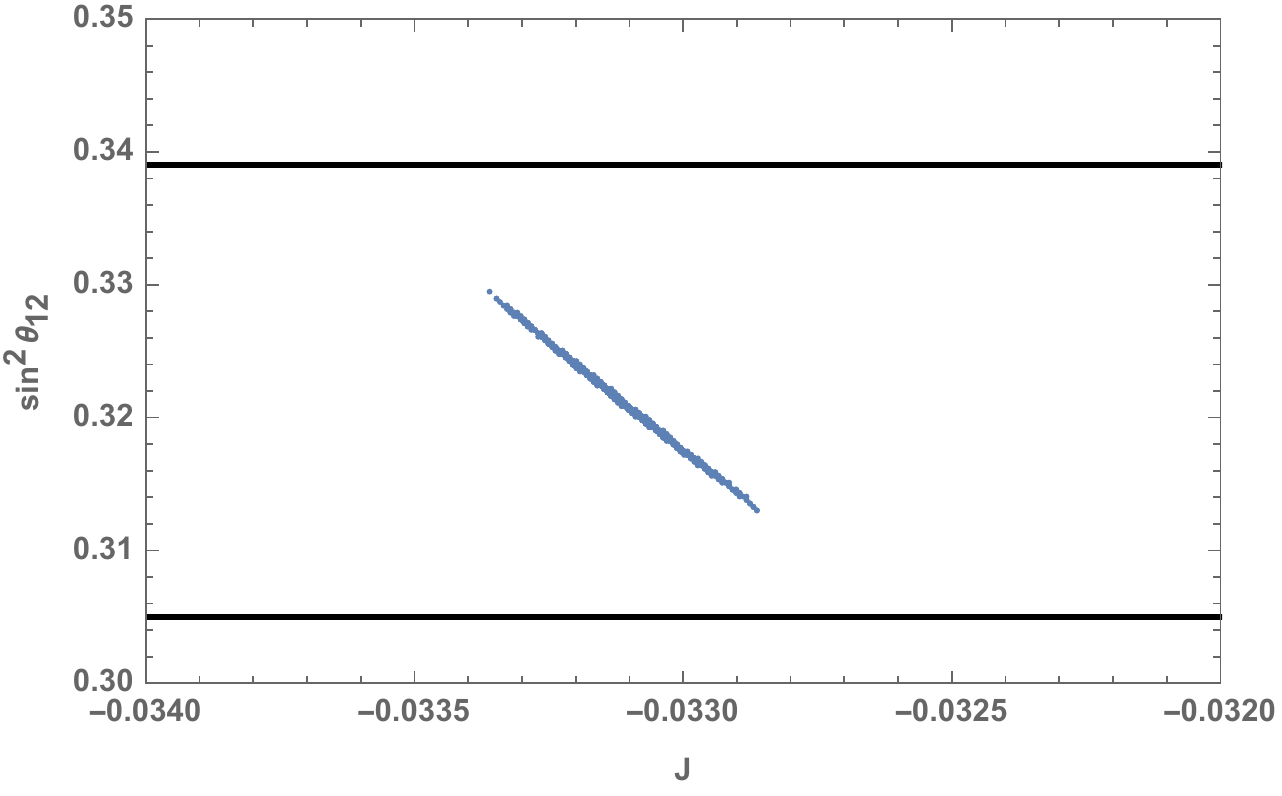}
\caption{Correlation between the solar mixing parameter $\sin
^{2}\theta _{12}$ and the Jarlskog invariant for the case of inverted neutrino mass hierarchy. The
horizontal lines are the minimum and maximum values of the solar mixing parameter $\sin^{2}\theta _{12}$ inside the $1\protect\sigma$ experimentally allowed range.}
\label{Correlation}
\end{figure}

As follows from Eqs.~(\ref{A1nu})-(\ref{f}), the obtained numerical values
given above for the neutrino parameters $\left\vert A_{1}^{\left( \nu
\right) }\right\vert $, $A_{2}^{\left( \nu \right) }$ and $\arg \left[
A_{1}^{\left( \nu \right) }\right] $ can be obtained from the following
benchmark point:
\begin{eqnarray}
m_{N_{1}} &=&500\hspace{0.5mm}\mbox{GeV},\hspace{0.5cm}m_{N_{2}}=2\hspace{%
0.5mm}\hspace{0.5mm}\mbox{TeV},\hspace{0.5cm}m_{\func{Re}\varphi }=900%
\hspace{0.5mm}\hspace{0.5mm}\mbox{GeV},\hspace{0.5cm}m_{\func{Im}\varphi
}=600\hspace{0.5mm}\mbox{GeV},  \notag \\
\Lambda &=&2.41\times 10^{5}~\mbox{TeV},\hspace{0.5cm}\left\vert y_{1\nu
}\right\vert =1.12,\hspace{0.5cm}y_{2\nu }=0.61,\hspace{0.5cm}\arg
\left[ y_{1\nu }\right] \simeq -37.4^{\circ }\,.
\label{benchmarkpointneutrino}
\end{eqnarray}
\begin{table}[t!]
\label{tab:leptons}
\begin{center}
\small
\begin{tabular}{|c||c|c|c|c|}
\hline
\multirow{2}{*}{Observable} & \multirow{2}{*}{Model value} & \multicolumn{3}{|c|}{Experimental value} \\\cline{3-5}
           &             &                          $1\sigma$ range & $2\sigma$ range & $3\sigma$ range \\ \hline\hline
$m_{e}$ [MeV] & $0.487$ & $0.487$ & $0.487$ & $0.487$\\ \hline
$m_{\mu }$ [MeV] & $102.8$ & $102.8\pm 0.0003$ & $102.8\pm0.0006$ & $102.8\pm 0.0009$ \\ \hline
$m_{\tau }$ [GeV] & $1.75$ & $1.75\pm 0.0003$ & $1.75\pm0.0006$ & $1.75\pm 0.0009$ \\ \hline
$\Delta m_{21}^{2}$ [$10^{-5}$eV$^{2}$] (IH) & $7.56$ & $7.56\pm0.19$ & $7.20-7.95$ & $7.05-8.14$ \\ \hline
$\Delta m_{13}^{2}$ [$10^{-3}$eV$^{2}$] (IH) & $2.49$ & $2.49\pm0.04$ & $2.41-2.57$ & $2.37-2.61$ \\ \hline

\multirow{2}{*}{$\delta $ [$^{\circ }$] (IH)} & \multirow{2}{*}{$281.6$} & \multirow{2}{*}{$259_{-41}^{+47}$} & \multirow{2}{*}{$182-347$} & $0-31$ \\
                             &         &                   &           & $142-360$ \\ \hline

$\sin ^{2}\theta _{12}$ (IH) & $0.321$ & $0.321_{-0.016}^{+0.018} $ & $0.289-0.359$ & $0.273-0.379$ \\ \hline
\multirow{2}{*}{$\sin ^{2}\theta _{23}$ (IH)} & \multirow{2}{*}{$0.511$} & \multirow{2}{*}{$0.596_{-0.018}^{+0.017}$} & $0.404-0.456$ & \multirow{2}{*}{$0.388-0.638$} \\ 
                             &               &                                  & $0.556-0.625$ &                     \\ \hline
$\sin ^{2}\theta _{13}$ (IH) & $0.0214$ & $0.0214_{-0.00085}^{+0.00082}$ & $0.0197-0.0230$ & $0.0189-0.0239$\\ \hline
\end{tabular}%
\end{center}
\par
\caption{Model and experimental values of the charged lepton masses,
neutrino mass squared splittings and leptonic mixing parameters for the
inverted (IH) mass hierarchy. The model values for CP violating phase are
also shown. The experimental values of the charged lepton masses are taken
from Ref.~\cite{Bora:2012tx}, whereas the range for experimental values
of neutrino mass squared splittings and leptonic mixing parameters, are
taken from Ref.~\cite{deSalas:2017kay}.}
\label{Observables0}
\end{table}
The benchmark point given above is one out of the many similar solutions that yields physical observables for the neutrino sector consistent with the experimental data. We have numerically checked that for a fixed mass splittings between the masses of the real and imaginary components of $\varphi$, the cutoff scale has a low sensitivity with the masses of the scalar and fermionic seesaw mediators. In addition, we have checked that lowering the mass splitting between $Re\varphi$ and $Im\varphi$ leads to a decrease of the cutoff scale. In particular lowering this mass splitting from $50\%$ up to $0.1\%$ of the mass of $Im\varphi$ leads to a decrease of the cutoff scale from $\sim 10^8$ GeV up to $\sim 10^7$ GeV. From Table \ref{Observables0}, it follows that the reactor $\sin ^{2}\theta _{13}$ and solar $\sin^{2}\theta _{12}$ leptonic mixing parameters are in excellent agreement with the experimental data, whereas the atmospheric $\sin ^{2}\theta _{23}$ mixing parameter is deviated $3\sigma $ away from its best fit value. Fig.~\ref{Correlation} shows the correlation between the solar mixing parameter $\sin^{2}\theta _{12}$ and the Jarlskog invariant for the case of inverted neutrino mass hierarchy. We found a leptonic Dirac CP violating phase of $281.6^{\circ }$ and a Jarlskog invariant close to about $-3.3\times 10^{-2}$ for the inverted neutrino mass hierarchy.

Let us consider the effective Majorana neutrino mass parameter
\begin{equation}
m_{ee}=\left\vert \sum_{j}U_{ek}^{2}m_{\nu _{k}}\right\vert ,  \label{mee}
\end{equation}
where $U_{ej}$ and $m_{\nu _{k}}$ are the PMNS leptonic
mixing matrix elements and the neutrino Majorana masses,
respectively. The neutrinoless double beta ($0\nu \beta \beta $) decay amplitude is proportional to $m_{ee}$. 
From Eq.~(\ref{eq:Mnu-1}) it follows that in our model there is a massless neutrino. 
It is well known that in this case, independently of the other parameters, one expects for the inverted neutrino mass hierarchy 
15 meV $< m_{ee}<$ 50 meV. With the model best fit values in Table~\ref{tab:leptons} we find
\begin{equation}
m_{ee}\simeq 41.3~\mbox{meV}\,.
\label{eff-mass-pred}
\end{equation}
This is within
the declared reach of the next-generation bolometric CUORE experiment \cite{Alessandria:2011rc} or, more realistically, of the
next-to-next-generation ton-scale $0\nu \beta \beta $-decay experiments. 
The current most stringent experimental upper limit
$m_{ee}\leq 160$ meV is set by 
$T_{1/2}^{0\nu
\beta \beta }(^{136}\mathrm{Xe})\geq 1.1\times 10^{26}$ yr at 90\% C.L.
from the KamLAND-Zen experiment \cite{KamLAND-Zen:2016pfg}. 

In theory, Lepton Flavour Violation processes are expected from this kind of model. However, in realisations such as these the new scale $\Lambda$ associated with family symmetry breaking scale is very high. Thus, the rate of muon conversion processes such as $\mu N \to e N$ ($N$ is nucleon), $\mu \to eee$, $\mu \to e \gamma$ is several orders of magnitude beyond experimental reach~\cite{Varzielas:2010mp}.

\section{Scalar Potential 
\label{scalar}}

In this section we consider the scalar potential. As can be seen in Table~\ref{ta:scalars}, the scalar content of the model has many degrees of
freedom. We assume that all scalars except for $\phi$ and $\varphi$ get their VEVs
at the family symmetry breaking scale, which should be near the cutoff scale 
$\Lambda$, much greater than the electroweak breaking scale defined by the
VEV of $\langle \phi \rangle \sim v$ (we can check the self-consistency of this assumption in the benchmark point in Eq.~(\ref{benchmarkpointneutrino})). Due to this, the family symmetry breaking
scalars decouple, such that we have at the TeV scale the effective potential 
$V(\phi,\varphi)$. We divide it into separate parts for convenience, and use
without loss of generality the mass eigenstates $\func{Re}\varphi$, $\func{Im%
}\varphi$ instead of $\varphi$, $\varphi^*$:
\begin{equation}
V(\phi ,\varphi )=V(\phi )+V(\phi ,\varphi )+V(\varphi )
\end{equation}%
where 
\begin{equation}
V(\phi )=-\mu ^{2}\left( \phi ^{\dagger }\cdot \phi \right) +\lambda \left(
\phi ^{\dagger }\cdot \phi \right) ^{2} + h.c,
\end{equation}%
is simply the SM potential (one Higgs doublet) and 
\begin{equation}
V(\phi ,\varphi )=\gamma _{1}\left( \phi ^{\dagger }\cdot \phi \right) \func{%
Re}\varphi ^{2}+\gamma _{2}\left( \phi ^{\dagger }\cdot \phi \right) \func{Im%
}\varphi ^{2} + h.c,
\end{equation}%
has only quartic interactions between the doublet $\phi $ and the $%
\mathbb{Z}_{2}$-odd scalar $\varphi $. The term 
\begin{equation}
V(\varphi )=-m_{1}^{2} \func{Re}\varphi ^2 - m_{2}^2 \func{Im}\varphi ^2 +
\lambda _{1} \left( \func{Re}\varphi \right) ^{4} + \lambda _{2} \left( 
\func{Im}\varphi \right) ^{4} + \lambda _{3} \left( \func{Re}\varphi^2 \func{%
Im}\varphi^2 \right) + h.c,
\end{equation}
has the masses and quartic interactions that involve only the $\mathbb{Z}_{2}
$ odd scalar. Given this, the masses of the real and imaginary parts of $%
\varphi$ will not be degenerate. As the symmetry is enhanced in the limit of
degeneracy (a $U(1)$ symmetry instead of the preserved $\mathbb{Z}_2$), if
the splitting between their masses is small it remains small, and a small
splitting is technically natural in that sense as it is protected by an
approximate symmetry.

\section{Dark Matter Constraints \label{DM}}

In this section we consider the possibilities offered by the model to
provide a viable DM candidate. The $\mathbb{Z}_{2}$ symmetry, under which only the
scalar field $\varphi $ and the fermions $N_{1R}$ and $N_{2R}$ are charged,
remains unbroken and stabilizes the lightest $\mathbb{Z}_{2}$-odd mass
eigenstate.

\subsection{Scalar Dark Matter Scenario}

\begin{figure}[t!]
\centering
\includegraphics[width=0.47\textwidth]{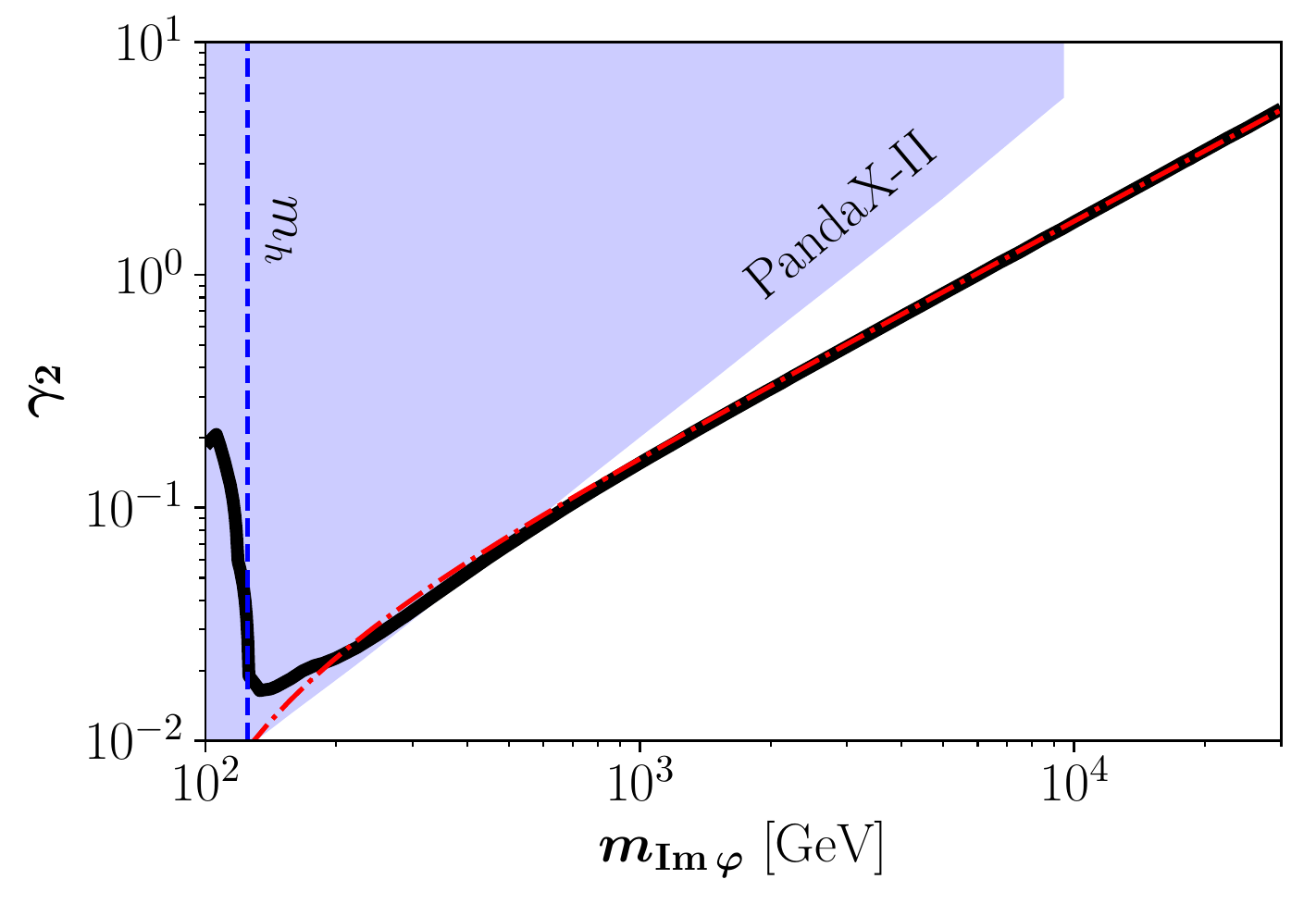}
\caption{Scalar Dark Matter Scenario. Parameter space generating the
observed DM relic abundance via the WIMP mechanism, using the full
annihilation cross-section (thick black line) and the only the annihilation
into Higgs bosons (thin red line). The light blue region is in tension with
the latest PandaX-II results.}
\label{fig:DM-Imphi}
\end{figure}

The first scenario considered is the one where one component of the scalar
field $\varphi$ is the lightest $\mathbb{Z}_2$-odd particle. In this case,
DM is produced in the early Universe via the vanilla WIMP paradigm. 
If Im~$\varphi$ is the lightest $\mathbb{Z}_2$ odd state, it can annihilate
into a pair of SM particles via the $s$-channel exchange of a Higgs boson.
Additionally, the annihilation into Higgs bosons also occurs via the contact
interaction and the mediation by an Im~$\varphi$ in the $t$- and $u$%
-channels. Finally, DM could also annihilate into a pair SM
neutrino/antineutrino via the $t$- and $u$-channel exchange of a $N_1$.
However the latter channel is typically very suppressed by the tiny
effective neutrino Yukawa coupling $y_{1\chi}\ll 1$. Hence, the DM relic
abundance is mainly governed by the DM mass $m_{\text{Im}~\varphi}$ and the
quartic coupling $\gamma_2$, between two DM particles and two Higgs bosons.
The freeze-out of heavy DM particles ($m_{\text{Im}~\varphi}> m_h$) is
largely dominated by the annihilations into Higgs bosons,\footnote{For $m_{\text{Im}~\varphi}=200$~GeV, annihilations into Higgses correspond to $\sim 80\%$ and into $t\bar t$ to $\sim 20\%$. When $m_{\text{Im}~\varphi}=10$~TeV, the annihilation into a pair of Higgses constitutes almost $100\%$.} with a
thermally-averaged cross-section given by: 
\begin{equation}  \label{DMhh}
\langle\sigma v\rangle\simeq\frac{\gamma_2^2}{32\pi}\,\left[\frac{%
\gamma_2\,v^2\,\left(m_h^2-4\,m_{\text{Im}~\varphi}^2\right)+m_h^4-4\,m_{%
\text{Im}~\varphi}^4}{m_{\text{Im}~\varphi}\,\left(m_h^4-6\,m_h^2\,m_{\text{%
Im}~\varphi}^2+8\,m_{\text{Im}~\varphi}^4\right)}\right]^2\,.
\end{equation}
In Fig.~\ref{fig:DM-Imphi} it is shown the parameter space ($\gamma_2,\,m_{%
\text{Im}~\varphi}$) giving rise to the observed DM relic abundance. The
black thick line corresponds to the full computation using micrOMEGAs~\cite%
{Belanger:2006is, Belanger:2008sj, Belanger:2010pz, Belanger:2013oya},
whereas the red line to the analytical case given by Eq.~\eqref{DMhh}. The
vertical dashed blue line corresponds to $m_{\text{Im}~\varphi}= m_h$.
The direct detection constraints are obtained by comparing the spin-independent cross section
for the scattering of the DM off of a nucleon,
\begin{equation}
\sigma_\text{SI}=\frac{\gamma_2^2\,m_N^4\,f^2}{8\pi\,m_h^4\,m_{\text{Im}~\varphi}^2}\,,
\end{equation}
to the latest limits on $\sigma_\text{SI}$ provided by PandaX-II~\cite{Cui:2017nnn}.
Here $m_N$ is the nucleon mass and $f\simeq 1/3$ corresponds to the form factor~\cite{Farina:2009ez, Giedt:2009mr}.
Again, the analytical result is in good agreement with the numerical computation by micrOMEGAs.
Fig.~\ref{fig:DM-Imphi} also presents the DM spin-independent direct detection exclusion region, that sets strong tension for the
model if the DM is lighter than $\sim 400$~GeV.\footnote{%
Furthermore, one has to take into account astrophysical uncertainties~\cite%
{Green:2002ht, Zemp:2008gw, McCabe:2010zh, Pato:2010yq, Fairbairn:2012zs,
Bernal:2014mmt, Bernal:2015oyn, Bernal:2016guq, Benito:2016kyp,
Green:2017odb} when interpreting the results of the DM searches.} 

\subsection{Fermionic Dark Matter Scenario}

The second case corresponds to the scenario where $N_{1R}$ is the lightest $%
\mathbb{Z}_{2}$-odd particle. DM can annihilate into a pair of SM neutrinos
via the $t$-channel exchange of the real and the imaginary parts of $\varphi$%
. This comes from an effective neutrino Yukawa coupling $y_{1\chi}\equiv%
\left\vert y_{1\nu }\right\vert\,\lambda\,\frac{v}{\Lambda}$ produced by
Eq.~(\ref{Lyn}) or its expanded version, Eq.~(\ref{eq:expanded_neutrinoL}):
\begin{equation}
\mathcal{L}\supset y_{1}^{\left( \nu \right) }\,\lambda\, \overline{l}%
_{1L}\,N_{1R}\,\varphi \frac{\langle \widetilde{\phi \rangle }}{\Lambda }\,.
\end{equation}%

\begin{figure}[t!]
\centering
\includegraphics[width=0.47\textwidth]{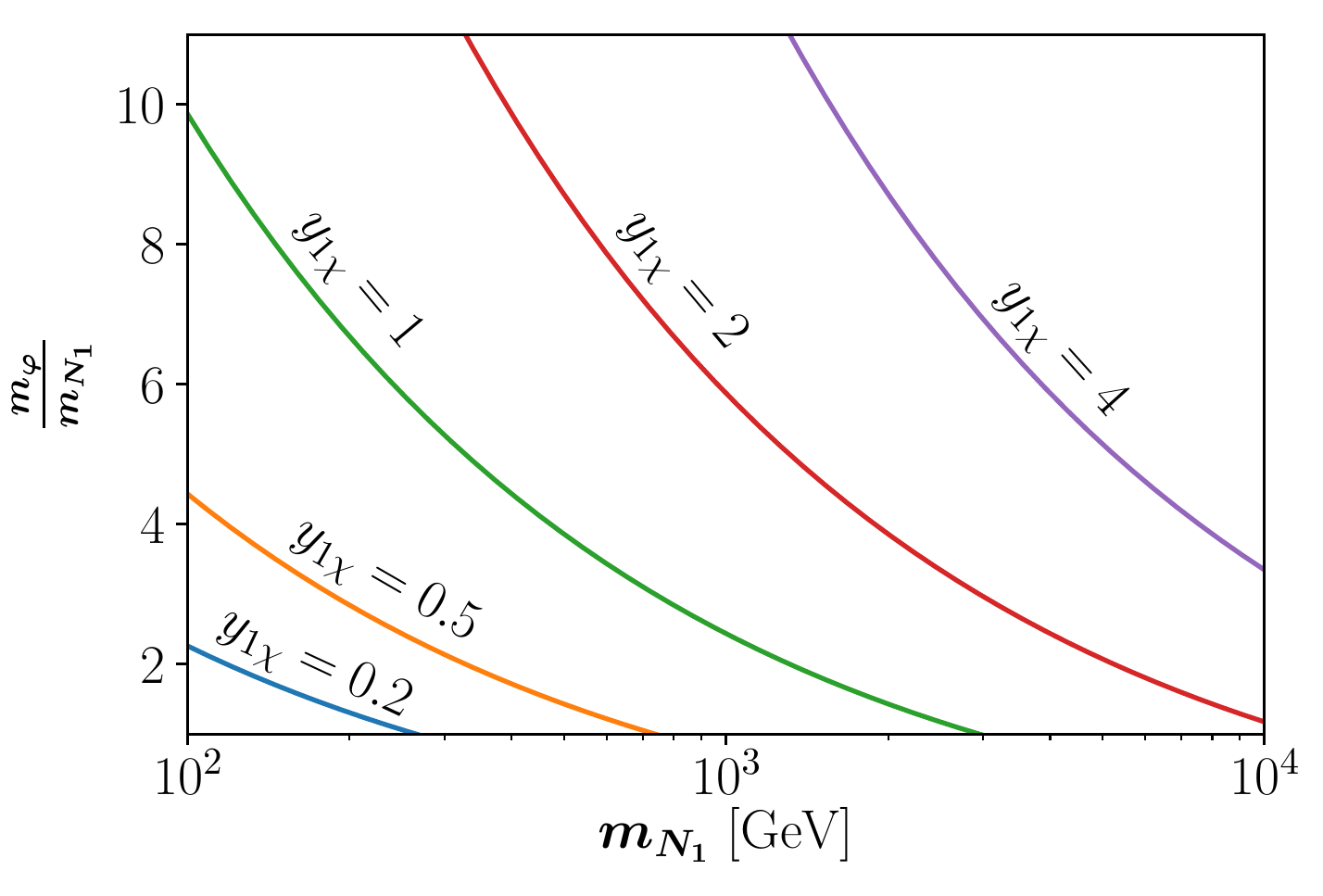} 
\caption{Fermionic Dark Matter Scenario. Effective coupling $y_{1\chi%
}$ needed in order to generate the observed relic abundance via the WIMP
mechanism, assuming $m_\varphi\equiv m_{\text{Re}\,\varphi}
\sim m_{\text{Im}\,\varphi}$.}
\label{fig:DM-masses}
\end{figure}

The DM relic abundance is then governed by the DM mass $m_{N_1}$, the
mediator masses $m_{\text{Re}\,\varphi}$ and $m_{\text{Im}\,\varphi}$, and
the effective Yukawa coupling $y_{1\chi}$. The thermally-averaged
annihilation cross-section is given by: 
\begin{equation}
\langle\sigma v\rangle\simeq\frac{9\,y_{1\chi}^4}{32\pi}\,\frac{%
m_{N_1}^2\,\left(2 m_{N_1}^2+m_{\text{Re}\,\varphi}^2+m_{\text{Im}%
\,\varphi}^2\right)^2}{\left(m_{N_1}^2+m_{\text{Re}\,\varphi}^2\right)^2\,%
\left(m_{N_1}^2+m_{\text{Im}\,\varphi}^2\right)^2}\,.
\end{equation}
Fig.~\ref{fig:DM-masses} shows the required effective coupling $y_{1\chi}$
in order to reproduce the observed DM relic abundance via the standard
thermal WIMP paradigm, and assuming $m_{\text{Re}\,\varphi} = m_{\text{Im}%
\,\varphi}$. 
As expected for WIMP DM, the effective coupling has to be of the order of
$\mathcal{O}(1)$, if DM is heavier than $\sim 100$~GeV. For the DM production this is
perfectly viable, however we also want to generate the neutrino masses. 
In what follows we proceed to scan for the CP odd scalar mass $m_{Im \varphi}
$ and effective neutrino Yukawa coupling $y_{1\chi}=\left\vert y_{1\nu
}\right\vert\,\lambda\,\frac{v}{\Lambda}$ needed required to reproduce the
values of the neutrino parameters $\left\vert A_{1}^{\left( \nu \right)
}\right\vert $, $A_{2}^{\left( \nu \right) }$ and $\arg \left[A_{1}^{\left(
\nu \right) }\right] $ shown in Eq.~\eqref{Fitpointneutrino}.
Fixing the right handed
Majorana neutrino masses to typical values $m_{N_1}\sim 500$~GeV, $m_{N_2}\sim 2$~TeV, $m_{\text{Im} \varphi}\sim 1$~TeV and $\Lambda\sim 10^{8}$~GeV,
required to reproduce the values of the neutrino
parameters $\left\vert A_{1}^{\left( \nu \right) }\right\vert $, $%
A_{2}^{\left( \nu \right) }$ and $\arg \left[A_{1}^{\left( %
\nu \right) }\right]$, 
the effective neutrino Yukawa coupling 
$y_{1\chi}$ has to be of the order of $10^{-7}$ to $10^{-4}$.
Values in this ballpark are too small to
reproduce the observed DM relic abundance via the WIMP mechanism, which
requires $\mathcal{O}(1)$ effective Yukawa coupling $y_{1\chi}$ as indicated
by Fig.~\ref{fig:DM-masses}. Consequently the fermionic DM scenario of our
model can not be produced via the usual WIMP paradigm.

\begin{figure}[t!]
\centering
\includegraphics[width=0.47\textwidth]{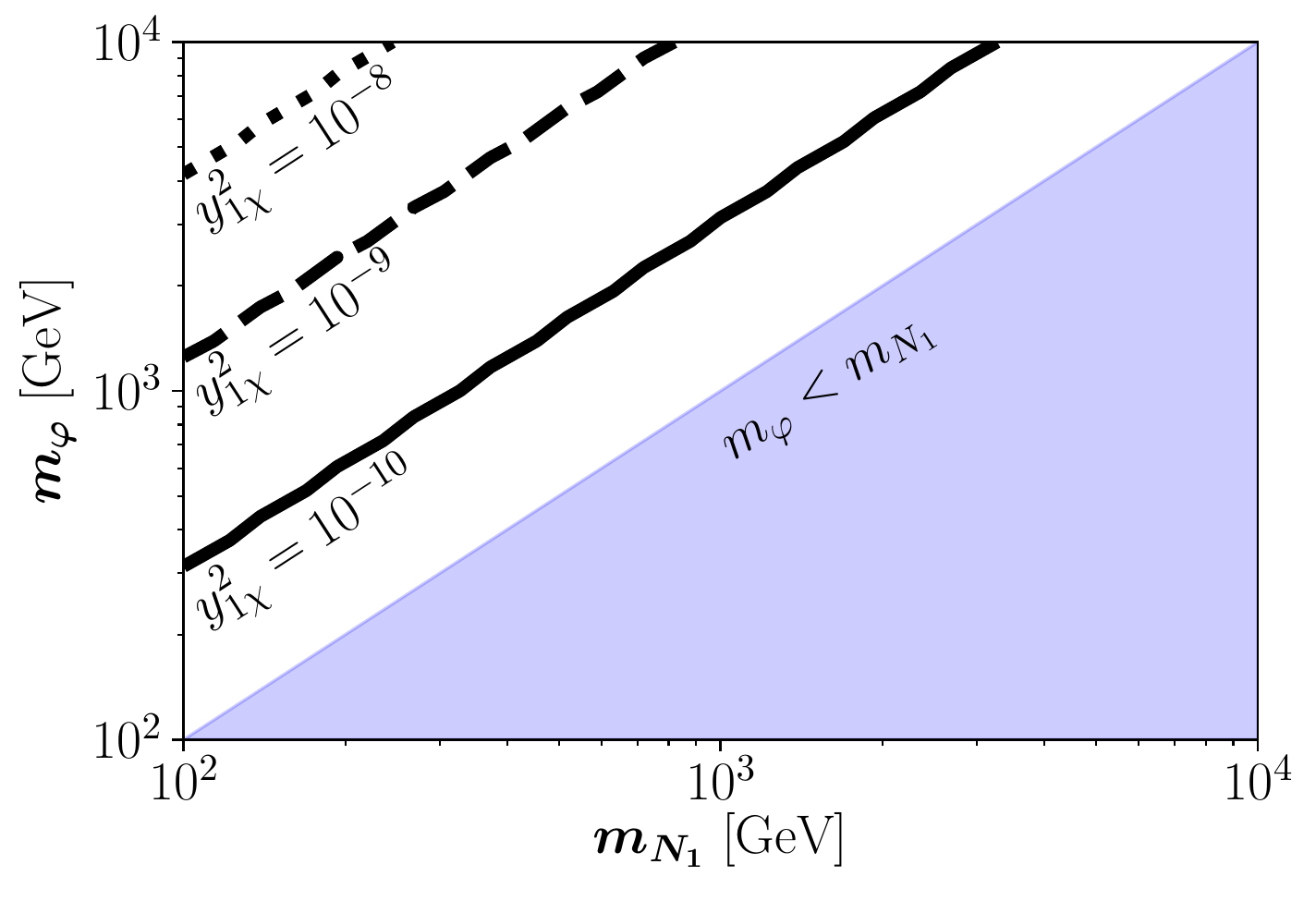} 
\caption{Fermionic Dark Matter Scenario. Effective coupling $y_{1\chi%
}^2$ needed in order to generate the observed relic abundance via the FIMP
mechanism, assuming $m_\varphi\equiv m_{\text{Re}\,\varphi}
\sim m_{\text{Im}\,\varphi}$.}
\label{fig:DM-masses2}
\end{figure}

Alternatively, very suppressed couplings between the visible and the dark sectors are characteristic in non-thermal scenarios where the DM relic abundance is created in the early Universe via freeze-in~\cite{McDonald:2001vt, Choi:2005vq, Kusenko:2006rh, Petraki:2007gq, Hall:2009bx, Bernal:2017kxu}.
Fig.~\ref{fig:DM-masses2} shows the effective couplings required in order to produce FIMP DM.
As expected for this kind of scenarios, $y_{1\chi}$ is in the range $\sim 10^{-8}$ to $\sim 10^{-11}$.
The light blue region is disregarded because $N_1$ is not the lightest particle of the dark sector.

Finally, to close this section, we discuss the splitting between the masses of the real and imaginary parts of $\varphi$.
To start, we note that a small scalar mass splitting of $10^{-3}$ times the mass of the imaginary part of $\varphi$ (which is required in order to have fermionic DM through the FIMP mechanism) may look unnatural,
but it is actually technically natural in the sense that it is protected by a symmetry: in the limit where the RH neutrino masses and the splitting of the $\varphi$ masses vanish, the symmetry of the Lagrangian is enlarged from the $\mathbb{Z}_2$ to a $U(1)$ symmetry. The non-trivial $U(1)$ charges of the RH neutrinos and of $\varphi$ under this $U(1)$ would forbid Majorana terms for the RH neutrinos and force the masses of the real and imaginary parts of $\varphi$ to be the same. Considering this, if the $U(1)$ is broken only by the Majorana terms (but not in the scalar potential), the splitting of the masses is no longer protected by the symmetry and is generated, but only radiatively. In such a scenario, the splitting would be naturally small.

Although we do not consider this scenario in great detail, we propose also some more explicit mechanisms that can
explain the splitting between the masses of the real and imaginary parts of $%
\varphi $ when starting from the symmetry limit where the splitting vanishes.
The first possibility consists in extending our model by adding an extra
spontaneously broken $\mathbb{Z}_{3}$ discrete symmetry under which $\varphi 
$ is assumed to have a charge $+1$ (in additive notation). In addition, an
extra SM scalar singlet, i.e. $\zeta $ , with $\mathbb{Z}_{3}$ charge +1
has to be added. The remaining scalar and fermions are neutral under $\mathbb{Z}_{3}$. Consequently no new contributions to the quarks, charged leptons and neutrino Yukawa terms originate from the extra field $\zeta$ and the $\mathbb{Z}_{3}$ discrete symmetry. The splitting between the masses of $\func{Re}\varphi $ and 
$\func{Im}\varphi $ will arise from the trilinear scalar interaction $%
A\varphi ^{2}\zeta $ which preserves both this added $\mathbb{Z}_{3}$ and the existing $\mathbb{Z}_{2}$.
The invariance of the neutrino Yukawa interactions
under the $\mathbb{Z}_{3}$ discrete symmetry requires that the right handed
Majorana neutrinos $N_{1R}$ and $N_{2R}$ should have a $\mathbb{Z}_{3}$
charge equal to $+1$, such that their masses will need to arise from the Yukawa interactions $%
\overline{N}_{1R}N_{1R}^{C}\zeta $ and $\overline{N}_{2R}N_{2R}^{C}\zeta $
after the spontaneous breaking of the $\mathbb{Z}_{3}$ discrete group. This is an explicit realization of the mechanism described above, showing there is a relation between the $\varphi$ mass splitting and the $N_{iR}$ masses. 
If this $\mathbb{Z}_{3}$ is broken at the TeV scale
the right handed Majorana neutrinos are within the LHC reach and there is a viable fermionic DM candidate through the FIMP mechanism.

A different mechanism to generate the splitting by
replacing the SM scalar singlet $\varphi $ with an inert $%
SU(2)$ scalar doublet charged under the preserved $\mathbb{Z}_{2}$
symmetry. That scenario was proposed for the first time in Ref.~\cite%
{Ma:2006km}. In that scenario, the splitting between the masses of $\func{Re}%
\varphi $ and $\func{Im}\varphi $ (in that scenario $\varphi $ is a $%
SU(2)$ scalar doublet) will arise form the quartic scalar
interaction $\left( \phi ^{\dagger }\cdot \varphi \right) ^{2}$, as
explained in detail in Ref.~\cite{Ma:2006km}. In this case, the coupling between right-handed neutrinos and $\varphi$ does not include the Higgs $\phi$.

\section{Conclusions}

We have built a viable family symmetry model based on the $\Delta \left( 27\right) \times \mathbb{Z}_{2}\times \mathbb{Z}_{5}\times \mathbb{Z}_{6}\times \mathbb{Z}_{10}\times \mathbb{Z}_{16}$ discrete group, which leads to a mixing inspired texture for the quarks and to similarly predictive structures for the leptons. For the quarks, the down sector parameters control the Cabibbo angle, and the up and down sector parameters control the remaining angles. For the leptons, the effective neutrino parameters that arise after radiative seesaw control the solar angle, and the charged lepton parameters control the reactor angle, which is also correlated to the deviation of the atmospheric angle from its maximal value. The model is only viable for inverted hierarchy and after fitting to the best-fit values of the solar and reactor angle, predicts $\sin^2 \theta_{23} \simeq 0.51$, $\delta \simeq 281.6^\circ$ and $m_{ee}=41.3$ meV.

Additionally, the model has viable DM candidates, stabilized by an unbroken $\mathbb{Z}_2$ symmetry, which we analyze quantitatively. A simple possibility is that there is scalar WIMP DM, which is produced through the Higgs portal. An alternative scenario is when we consider fermionic DM, which in our model would be the lightest right-handed neutrino. In order for it to be a WIMP and to obtain the right abundance, its effective coupling to the visible sector is too large to be consistent with what is required by the effective neutrino masses. Instead, if our fermionic DM candidate is a FIMP, the effective coupling needs to be quite small. This is consistent with obtaining the required neutrino masses but requires a very small splitting of the real and imaginary components of the $\mathbb{Z}_2$-odd scalar (the splitting divided by the mass scale would be at the per mille level). The smallness of the splitting is technically natural as when the splitting goes to zero, the symmetry of the theory is enhanced.

This model addresses the flavour problem while providing a viable DM candidate (scalar or fermionic), and is a novel example of the interplay of constraints coming from the observed DM abundance to a family symmetry model, namely by relating the DM abundance to the light neutrino masses.


\section*{Acknowledgments}
IdMV acknowledges funding from the Funda\c{c}\~{a}o para a Ci\^{e}ncia e a Tecnologia (FCT) through
the contract IF/00816/2015, partial support by Funda\c{c}\~ao para a Ci\^encia e a Tecnologia
(FCT, Portugal) through the project CFTP-FCT Unit 777 (UID/FIS/00777/2013) which
is partially funded through POCTI (FEDER), COMPETE, QREN and EU,
and partial support by the National Science Center, Poland, through the HARMONIA project under contract UMO-2015/18/M/ST2/00518. IdMV thanks Universidad Técnica Federico Santa
Mar\'{\i}a for hospitality, where this work was finished. The visit of IdMV to Universidad T\'{e}cnica Federico Santa Mar\'{\i}a was supported by Chilean grant Fondecyt No. 1170803. NB is partially supported by the Spanish MINECO under Grants FPA2014-54459-P and FPA2017-84543-P.
This project has received funding from the European Union's Horizon 2020 research and innovation programme under the Marie Skłodowska-Curie grant agreements 674896 and 690575; and from Universidad Antonio Nariño grant 2017239. AECH and SK were supported by Chilean grants Fondecyt No. 1170803, No. 1150792 and CONICYT
PIA/Basal FB0821, ACT1406 and the UTFSM internal grant PI\_M\_17\_5. A.E.C.H is very grateful to the Instituto Superior T\'{e}cnico for hospitality. 

\appendix

\section{The Product Rules of the $\boldsymbol{\Delta (27)}$ Discrete Group}

\label{A}

The $\Delta (27)$ discrete group is a subgroup of $SU(3)$, has 27 elements
divided into 11 conjugacy classes. Then the $\Delta (27)$ discrete group
contains the following 11 irreducible representations: two triplets, i.e. $%
\mathbf{3}_{[0][1]}$ (which we denote by $\mathbf{3}$)\ and its conjugate $%
\mathbf{3}_{[0][2]}$ (which we denote by $\overline{\mathbf{3}}$) and 9
singlets, i.e. $\mathbf{1}_{k,l}$ ($k,l=0,1,2$), where $k$ and $l$
correspond to the $\mathbb{Z}_{3}$ and $\mathbb{Z}_{3}^{\prime }$ charges,
respectively~\cite{Ishimori:2010au}. The $\Delta (27)$ discrete group, which
is a simple group of the type $\Delta (3n^{2})$ with $n=3$, is isomorphic to
the semi-direct product group $(\mathbb{Z}_{3}^{\prime }\times \mathbb{Z}%
_{3}^{\prime \prime })\rtimes \mathbb{Z}_{3}$~\cite{Ishimori:2010au}. It is
worth mentioning that the simplest group of the type $\Delta (3n^{2})$ is $%
\Delta (3)\equiv \mathbb{Z}_{3}$. The next group is $\Delta (12)$, which is
isomorphic to $A_{4}$. Consequently the $\Delta (27)$ discrete group is the
simplest nontrivial group of the type $\Delta(3n^{2})$. Any element of the $%
\Delta (27)$ discrete group can be expressed as $b^{k}a^{m}{a^{\prime }}^{n}$%
, being $b$, $a$ and $a^{\prime }$ the generators of the $\mathbb{Z}_{3}$, $%
\mathbb{Z}_{3}^{\prime }$ and $\mathbb{Z}_{3}^{\prime \prime }$ cyclic
groups, respectively. These generators fulfill the relations: 
\begin{eqnarray}
&& a^3 =a^{\prime 3}=b^3=1,\hspace*{0.5cm} a a^{\prime }=a^{\prime }a, 
\notag \\
&& bab^{-1}=a^{-1}a^{\prime -1},\,\, b a^{\prime}b^{-1}=a.  \label{aapbrela}
\end{eqnarray}%
The characters of the $\Delta (27)$ discrete group are shown in Table~\ref%
{tab:delta-27}. Here $n$ is the number of elements, $h$ is the order of each
element, and $\omega =e^{\frac{2\pi i}{3}}=-\frac{1}{2}+i\frac{\sqrt{3}}{2}$
is the cube root of unity, which satisfies the relations $1+\omega +\omega
^{2}=0$ and $\omega ^{3}=1$. 
The conjugacy classes of $\Delta (27)$ are given by: 
\begin{equation*}
\begin{array}{ccc}
C_{1}: & \{e\}, & h=1, \\ 
C_{1}^{(1)}: & \{a,a^{\prime 2}\}, & h=3, \\ 
C_{1}^{(2)}: & \{a^{2},a^{\prime }\}, & h=3, \\ 
C_{3}^{(0,1)}: & \{a^{\prime 2}a^{\prime 2}\}, & h=3, \\ 
C_{3}^{(0,2)}: & \{a^{\prime 2},a^{2},aa^{\prime }\}, & h=3, \\ 
C_{3}^{(1,p)}: & \{ba^{p},ba^{p-1}a^{\prime p-2}a^{\prime 2}\}, & h=3, \\ 
C_{3}^{(2,p)}: & \{ba^{p},ba^{p-1}a^{\prime p-2}a^{\prime 2}\}, & h=3. \\ 
&  & 
\end{array}%
\end{equation*}
\begin{table}[t]
\begin{center}
\begin{tabular}{|c||c|c|c|c|}
\hline
& h & $\chi_{1_{(r,s)}}$ & $\chi_{3_{[0,1]}}$ & $\chi_{3_{[0,2]}}$ \\ \hline\hline
$1C_1$ & 1 & 1 & 3 & 3 \\ \hline
$1C_1^{(1)}$ & 1 & 1 & $3\omega^2$ & $3\omega$ \\ \hline
$1C_1^{(2)}$ & 1 & 1 & $3\omega$ & $3\omega^2$ \\ \hline
$3C_1^{(0,1)}$ & $3$ & $\omega^{s}$ & $0$ & $0$ \\ \hline
$3C_1^{(0,2)}$ & $3$ & $\omega^{2s}$ & $0$ & $0$ \\ \hline
$C_3^{(1,p)}$ & $3$ & $\omega^{r+s p}$ & 0 & 0 \\ \hline
$C_3^{(2,p)}$ & $3$ & $\omega^{2r+s p}$ & 0 & 0 \\ \hline
\end{tabular}%
\end{center}
\caption{Characters of $\Delta (27)$.}
\label{tab:delta-27}
\end{table}
The tensor products between $\Delta (27)$ triplets are described by the
following relations~\cite{Ishimori:2010au}: 
\begin{eqnarray}
\vspace{-1cm}%
\begin{pmatrix}
x_{1,-1} \\ 
x_{0,1} \\ 
x_{-1,0} \\ 
\end{pmatrix}%
_{\mathbf{3}_{[0][1]}}\otimes 
\begin{pmatrix}
y_{1,-1} \\ 
y_{0,1} \\ 
y_{-1,0} \\ 
\end{pmatrix}%
_{\mathbf{3}_{[0][1]}} &=&%
\begin{pmatrix}
x_{1,-1}y_{1,-1} \\ 
x_{0,1}y_{0,1} \\ 
x_{-1,0}y_{-1,0} \\ 
\end{pmatrix}%
_{\mathbf{3}_{[0][2]}^{\left( S_{1}\right) }}\oplus \frac{1}{2}%
\begin{pmatrix}
x_{0,1}y_{-1,0}+x_{-1,0}y_{0,1} \\ 
x_{-1,0}y_{1,-1}+x_{1,-1}y_{-1,0} \\ 
x_{1,-1}y_{0,1}+x_{0,1}y_{1,-1} \\ 
\end{pmatrix}%
_{\mathbf{3}_{[0][2]}^{\left( S_{2}\right) }}  \notag \\
&&\oplus \frac{1}{2}%
\begin{pmatrix}
x_{0,1}y_{-1,0}-x_{-1,0}y_{0,1} \\ 
x_{-1,0}y_{1,-1}-x_{1,-1}y_{-1,0} \\ 
x_{1,-1}y_{0,1}-x_{0,1}y_{1,-1} \\ 
\end{pmatrix}%
_{\mathbf{3}_{[0][2]}^{\left( A\right) }}, \\
\begin{pmatrix}
x_{2,-2} \\ 
x_{0,2} \\ 
x_{-2,0} \\ 
\end{pmatrix}%
_{\mathbf{3}_{[0][2]}}\otimes 
\begin{pmatrix}
y_{2,-2} \\ 
y_{0,2} \\ 
y_{-2,0} \\ 
\end{pmatrix}%
_{\mathbf{3}_{[0][2]}} &=&%
\begin{pmatrix}
x_{2,-2}y_{2,-2} \\ 
x_{0,2}y_{0,2} \\ 
x_{-2,0}y_{-2,0} \\ 
\end{pmatrix}%
_{\mathbf{3}_{[0][1]}^{\left( S_{1}\right) }}\oplus \frac{1}{2}%
\begin{pmatrix}
x_{0,2}y_{-2,0}+x_{-2,0}y_{0,2} \\ 
x_{-2,0}y_{2,-2}+x_{2,-2}y_{-2,0} \\ 
x_{2,-2}y_{0,2}+x_{0,2}y_{2,-2} \\ 
\end{pmatrix}%
_{\mathbf{3}_{[0][1]}^{\left( S_{2}\right) }}  \notag \\
&&\oplus \frac{1}{2}%
\begin{pmatrix}
x_{0,2}y_{-2,0}-x_{-2,0}y_{0,2} \\ 
x_{-2,0}y_{2,-2}-x_{2,-2}y_{-2,0} \\ 
x_{2,-2}y_{0,2}-x_{0,2}y_{2,-2} \\ 
\end{pmatrix}%
_{\mathbf{3}_{[0][1]}^{\left( A\right) }}, \\
\begin{pmatrix}
x_{1,-1} \\ 
x_{0,1} \\ 
x_{-1,0} \\ 
\end{pmatrix}%
_{\mathbf{3}_{[0][1]}}\otimes 
\begin{pmatrix}
y_{-1,1} \\ 
y_{0,-1} \\ 
y_{1,0} \\ 
\end{pmatrix}%
_{\mathbf{3}_{[0][2]}} &=&\sum_{r}(x_{1,-1}y_{-1,1}+\omega
^{2r}x_{0,1}y_{0,-1}+\omega ^{r}x_{-1,0}y_{1,0})_{\mathbf{1}_{(r,0)}}  \notag
\\
&\oplus &\sum_{r}(x_{1,-1}y_{0,-1}+\omega ^{2r}x_{0,1}y_{1,0}+\omega
^{r}x_{-1,0}y_{-1,1})_{\mathbf{1}_{(r,1)}}  \notag \\
&\oplus &\sum_{r}(x_{1,-1}y_{1,0}+\omega ^{2r}x_{0,1}y_{-1,1}+\omega
^{r}x_{-1,0}y_{0,-1})_{\mathbf{1}_{(r,2)}}.  \notag \\
&&
\end{eqnarray}%
The multiplication rules between $\Delta (27)$ singlets and $\Delta (27)$
triplets are given by~\cite{Ishimori:2010au}: 
\begin{eqnarray}
&&%
\begin{pmatrix}
x_{(1,-1)} \\ 
x_{(0,1)} \\ 
x_{(-1,0)}%
\end{pmatrix}%
_{\mathbf{3}_{[0][1]}}\otimes (z)_{1_{k,l}}=%
\begin{pmatrix}
x_{(1,-1)}z \\ 
\omega ^{r}x_{(0,1)}z \\ 
\omega ^{2r}x_{(-1,0)}z%
\end{pmatrix}%
_{\mathbf{3}_{[l][1+l]}}, \\
&&%
\begin{pmatrix}
x_{(2,-2)} \\ 
x_{(0,2)} \\ 
x_{(-2,0)}%
\end{pmatrix}%
_{\mathbf{3}_{[0][2]}}\otimes (z)_{1_{k,l}}=%
\begin{pmatrix}
x_{(2,-2)}z \\ 
\omega ^{r}x_{(0,2)}z \\ 
\omega ^{2r}x_{(-2,0)}%
\end{pmatrix}%
_{\mathbf{3}_{[l][2+l]}}.
\end{eqnarray}%
The tensor products of $\Delta (27)$ singlets $\mathbf{1}_{k,\ell }$ and $%
\mathbf{1}_{k^{\prime },\ell ^{\prime }}$ take the form~\cite%
{Ishimori:2010au}: 
\begin{equation}
\mathbf{1}_{k,\ell }\otimes \mathbf{1}_{k^{\prime },\ell ^{\prime }}=\mathbf{%
1}_{k+k^{\prime }\func{mod}3,\ell +\ell ^{\prime }\func{mod}3}.
\end{equation}%
From the equation given above, we obtain explicitly the singlet
multiplication rules of the $\Delta (27)$ group, which are given in Table~\ref{D27multiplets}. 
\begin{table}[t!]
\begin{center}
\begin{tabular}{|c||c|c|c|c|c|c|c|c|}
\hline
Singlets & ~ $\mathbf{1}_{01}$ ~ & ~ $\mathbf{1}_{02}$ ~ & ~ $\mathbf{1}%
_{10} $ ~ & ~ $\mathbf{1}_{11}$~ & ~ $\mathbf{1}_{12}$ ~ & ~ $\mathbf{1}%
_{20} $ ~ & ~ $\mathbf{1}_{21}$~ & $\mathbf{1}_{22}$ \\ \hline\hline
$\mathbf{1}_{01}$ & $\mathbf{1}_{02}$ & $\mathbf{1}_{00}$ & $\mathbf{1}_{11}$
& $\mathbf{1}_{12}$ & $\mathbf{1}_{10}$ & $\mathbf{1}_{21}$ & $\mathbf{1}%
_{22}$ & $\mathbf{1}_{20}$ \\ \hline
$\mathbf{1}_{02}$ & $\mathbf{1}_{00}$ & $\mathbf{1}_{01}$ & $\mathbf{1}_{12}$
& $\mathbf{1}_{10}$ & $\mathbf{1}_{11}$ & $\mathbf{1}_{22}$ & $\mathbf{1}%
_{20}$ & $\mathbf{1}_{21}$ \\ \hline
$\mathbf{1}_{10}$ & $\mathbf{1}_{11}$ & $\mathbf{1}_{12}$ & $\mathbf{1}_{20}$
& $\mathbf{1}_{21}$ & $\mathbf{1}_{22}$ & $\mathbf{1}_{00}$ & $\mathbf{1}%
_{01}$ & $\mathbf{1}_{02}$ \\ \hline
$\mathbf{1}_{11}$ & $\mathbf{1}_{12}$ & $\mathbf{1}_{10}$ & $\mathbf{1}_{21}$
& $\mathbf{1}_{22}$ & $\mathbf{1}_{20}$ & $\mathbf{1}_{01}$ & $\mathbf{1}%
_{02}$ & $\mathbf{1}_{00}$ \\ \hline
$\mathbf{1}_{12}$ & $\mathbf{1}_{10}$ & $\mathbf{1}_{11}$ & $\mathbf{1}_{22}$
& $\mathbf{1}_{20}$ & $\mathbf{1}_{21}$ & $\mathbf{1}_{02}$ & $\mathbf{1}%
_{00}$ & $\mathbf{1}_{01}$ \\ \hline
$\mathbf{1}_{20}$ & $\mathbf{1}_{21}$ & $\mathbf{1}_{22}$ & $\mathbf{1}_{00}$
& $\mathbf{1}_{01}$ & $\mathbf{1}_{02}$ & $\mathbf{1}_{10}$ & $\mathbf{1}%
_{11}$ & $\mathbf{1}_{12}$ \\ \hline
$\mathbf{1}_{21}$ & $\mathbf{1}_{22}$ & $\mathbf{1}_{20}$ & $\mathbf{1}_{01}$
& $\mathbf{1}_{02}$ & $\mathbf{1}_{00}$ & $\mathbf{1}_{11}$ & $\mathbf{1}%
_{12}$ & $\mathbf{1}_{10}$ \\ \hline
$\mathbf{1}_{22}$ & $\mathbf{1}_{20}$ & $\mathbf{1}_{21}$ & $\mathbf{1}_{02}$
& $\mathbf{1}_{00}$ & $\mathbf{1}_{01}$ & $\mathbf{1}_{12}$ & $\mathbf{1}%
_{10}$ & $\mathbf{1}_{11}$ \\ \hline
\end{tabular}%
\end{center}
\caption{The singlet multiplications of the group $\Delta (27)$.}
\label{D27multiplets}
\end{table}

\bibliographystyle{JHEP}
\bibliography{biblio}


\end{document}